\documentclass[12pt]{article} 
\pdfoutput=1

\usepackage{amsmath,amssymb,amsfonts}
\usepackage{color}
\definecolor{darkblue}{rgb}{0.1,0.1,.7}
\usepackage[colorlinks, linkcolor=darkblue, citecolor=darkblue, urlcolor=darkblue, linktocpage]{hyperref} 
\usepackage[square, comma, compress,numbers]{natbib}
\usepackage[]{amsmath}
\usepackage[]{graphicx}
\usepackage[]{latexsym}
\usepackage{dsfont}
\usepackage{geometry}
\usepackage{amscd}
\usepackage[all,cmtip]{xy}
\usepackage{mathrsfs}
\usepackage{bbold}
\usepackage{subfigure}
\usepackage[margin=10pt,font=small,labelfont=bf]{caption}
\usepackage{dsdshorthand}
\usepackage{simplewick}
\usepackage{changepage}
\usepackage[]{algorithm2e}
\usepackage{booktabs,multirow}
\newenvironment{lyxcode}
	{\par\begin{list}{}{
		\setlength{\rightmargin}{\leftmargin}
		\setlength{\listparindent}{0pt}
		\raggedright
		\setlength{\itemsep}{0pt}
		\setlength{\parsep}{0pt}
		\normalfont\ttfamily}%
	 \item[]}
	{\end{list}}

\newcommand{\es}[2] {\begin{equation} \label{#1} \begin{split} #2 \end{split} \end{equation}}

\newcommand\Otwo{${O}(2)$ }
\newcommand\Othree{${O}(3)$ }
\newcommand\ON{${O}(N)$ }
\newcommand\Othreewithout{{O}(3)}
\newcommand\Otwowithout{{O}(2)}
\newcommand\SOthree{${SO}(3)$ }
\newcommand\SOthreewithout{{SO}(3) }

\newcommand\Ononewithout{{O}(n_1)}
\newcommand\Ontwowithout{{O}(n_2)}
\newcommand\Ononetwowithout{{O}(n_1+n_2)}
\newcommand\Ononetwodirectwithout{{O}(n_1) \times {O}(n_2)}

\begin{document}

\vspace*{-.6in} \thispagestyle{empty}
\begin{flushright}
CALT-TH-2020-053
\end{flushright}
\vspace{.2in} {\Large
\begin{center}
{\bf 
Bootstrapping Heisenberg Magnets \\and their Cubic Instability
}
\end{center}
}
\vspace{.2in}
\begin{center}
{\bf 
Shai M. Chester$^{a}$,
Walter Landry$^{b,c}$,
Junyu Liu$^{c,d}$,
David Poland$^{e}$,\\
David Simmons-Duffin$^{c}$,
Ning Su$^{f}$,
Alessandro Vichi$^{f,g}$} 
\\
\vspace{.2in} 
$^a$ {\it Department of Particle Physics and Astrophysics, Weizmann Institute of Science, Rehovot, Israel}\\
$^b$ {\it Simons Collaboration on the Nonperturbative Bootstrap}\\
$^c$ {\it Walter Burke Institute for Theoretical Physics, Caltech, Pasadena, CA 91125, USA}\\
$^d$ {\it Institute for Quantum Information and Matter, Caltech, Pasadena, CA 91125, USA}\\
$^e$ {\it Department of Physics, Yale University, New Haven, CT 06520, USA}\\
$^f$ {\it Institute of Physics,
\'Ecole Polytechnique F\'ed\'erale de Lausanne (EPFL),\\
 CH-1015 Lausanne, Switzerland}\\
 $^g$ {\it Department of Physics, University of Pisa, I-56127 Pisa, Italy}

\end{center}

\vspace{.2in}

\begin{abstract}

We study the critical $O(3)$ model using the numerical conformal bootstrap. In particular, we use a recently developed cutting-surface algorithm to efficiently map out the allowed space of CFT data from correlators involving the leading $O(3)$ singlet $s$, vector $\phi$, and rank-2 symmetric tensor $t$. We determine their scaling dimensions to be $(\Delta_{s}, \Delta_{\phi}, \Delta_{t}) = (0.518942(51), 1.59489(59), 1.20954(23))$, and also bound various OPE coefficients. We additionally introduce a new ``tip-finding" algorithm to compute an upper bound on the leading rank-4 symmetric tensor $t_4$, which we find to be relevant with $\Delta_{t_4} < 2.99056$. The conformal bootstrap thus provides a numerical proof that systems described by the critical $O(3)$ model, such as classical Heisenberg ferromagnets at the Curie transition, are unstable to cubic anisotropy.

\end{abstract}

\newpage

\renewcommand{\baselinestretch}{0.95}\normalsize
\tableofcontents
\renewcommand{\baselinestretch}{1.0}\normalsize

\newpage

\section{Introduction}

Numerical bootstrap methods \cite{Rattazzi:2008pe,Rychkov:2009ij} (see \cite{Poland:2018epd,Chester:2019wfx} for recent reviews) have led to powerful new results in the study of conformal field theories (CFTs). In \cite{Chester:2019ifh,Liu:2020tpf} we developed an approach to large-scale bootstrap problems which allowed for precise determinations of the CFT data of the 3d critical \Otwo model. In this work, we continue the exploration of large-scale bootstrap problems by applying the technology introduced in \cite{Chester:2019ifh} to the study of the 3d critical $O(3)$ model.

Concretely, we apply these methods to study correlation functions of the lowest-dimension singlet, vector, and rank-2 scalars in the three-dimensional critical $O(3)$ model.
Using the ``cutting surface" algorithm introduced in~\cite{Chester:2019ifh}, 
we compute the allowed region for the CFT data of these leading scalar operators. Our results, together with comparisons to results from Monte Carlo simulations, are summarized in table~\ref{tab:results}. We also introduce a new algorithm and software implementation called \texttt{tiptop}, which allows us to efficiently test allowed gaps for other operators across this region. We use it to determine an upper bound on the dimension of the lowest-dimension rank-4 scalar.

The 3d $O(3)$ model is a well-studied renormalization group (RG) fixed point, and its critical exponents have been computed using many methods, both theoretical and experimental. This model describes the critical behavior of isotropic magnets, such as the Curie transition in isotropic ferromagnets, and antiferromagnets at the N\'eel transition point. Moreover, since disorder corresponds to an irrelevant perturbation,\footnote{This is the case in any \ON model with $N\geq2$.} the model also describes isotropic magnets with quenched disorder.
 
One of the main open questions about the $O(3)$ model is its stability under cubic deformations. The majority of magnets present in nature are indeed not isotropic: this means that the microscopic Hamiltonian describing the system in the ultraviolet (UV) is not invariant under the full $O(3)$ symmetry group but only under a discrete subgroup, such as the cubic symmetry group. This implies that additional terms will be generated at the microscopic level that are invariant under cubic symmetry but transform in a non-trivial representation of $O(3)$. If any of those deformations turn out to be relevant, the $O(3)$ fixed point would be unstable and could not be reached without further tuning in the UV theory. The attractive, stable, fixed point would instead be the 3d cubic model. Field theory computations and Monte Carlo simulations have shown that these two models have very similar critical exponents: hence, if the cubic perturbation is relevant, it should be very close to marginality and the RG flow connecting the two theories is very short. We will come back to this point in section~\ref{sec:ONvscubic}.

\begin{table*}[ht]
\centering
\begin{tabular}{@{}cc|cc@{}}
	\toprule
CFT data & method & value & ref \\
	\midrule
$\Delta_s$ & MC & 1.5948(2) &\cite{Hasenbusch:2020pwj} \\ 
& CB & 1.5957({\bf 55}) & \cite{Kos:2016ysd} \\
 & CB & 1.59488({\bf 81}) & \text{this work}\\
	 \midrule
 $\Delta_\phi$ & MC & 0.518920(25) & \cite{Hasenbusch:2020pwj} \\
 & CB & 0.51928({\bf 62}) & \cite{Kos:2016ysd} \\
 & CB & 0.518936({\bf 67}) & \text{this work}\\
	 \midrule
 $\Delta_t$ & MC & 1.2094(3) & \cite{Hasenbusch_2011} \\
 & CB & $1.2095(55^*)$& \cite{Kos:2015mba} \\
 & CB & 1.20954({\bf 32}) & \text{this work}\\
 \midrule
$\Delta_{t_4}$ & MC & 2.987(4) & \cite{Hasenbusch_2011} \\
 & CB & $< 2.99056$ & \text{this work}\\
 \midrule
$\lambda_{\f\f s}$  & CB & $0.5244(11^*)$ & \cite{Kos:2016ysd} \\
& CB & $0.524261(59^*)$ & \text{this work}\\
\midrule
$\lambda_{sss}$ & CB & $0.499(12^*)$ & \cite{Kos:2016ysd} \\
& CB & $0.5055(11^*)$ & \text{this work} \\
\midrule
$\lambda_{tts}$ & CB & $0.98348(39^*)$ & \text{this work} \\
\midrule
$\lambda_{\f\f t}$ & CB & $0.87451(22^*)$ & \text{this work}\\
\midrule
$\lambda_{ttt}$ & CB & $1.49957(49^*)$ & \text{this work}\\
\bottomrule 
\end{tabular}
	\caption{Comparison of conformal bootstrap (CB) results with previous determinations from Monte Carlo (MC) simulations. We denote the leading rank-0, rank-1, rank-2, and rank-4 scalars by $s,\f,t,t_4$, respectively. Bold uncertainties correspond to rigorous intervals from bootstrap bounds. Uncertainties marked with a $^*$ indicate that the value is estimated non-rigorously by sampling points. 
	\label{tab:results}}
\end{table*}

We give a definite answer to the above question: the $O(3)$ model is unstable under cubic deformations. This information is encoded in the dimension of the lowest rank-4 scalar ${t_4}$, which in the $O(3)$ model satisfies $\Delta_{t_4}<3$. As we will discuss, this implies that the $O(3)$ model is also unstable with respect to the biconal fixed point with $\mathbb{Z}_2 \times O(2)$ symmetry. Relevance of $t_4$ has been previously suggested by Monte Carlo~\cite{Hasenbusch_2011} and perturbative expansions~\cite{Carmona:1999rm}, but the proximity to marginality and near degeneracy of the critical exponents between the cubic, biconal, and $O(3)$ fixed points makes this a subtle question ideal for the precision and rigor of the conformal bootstrap.

\subsection{Theoretical approaches to the 3d $O(3)$ model}

We start by briefly reviewing past approaches to the 3d $O(3)$ model, including field theory studies, Monte Carlo, and past results obtained by conformal bootstrap techniques. We also describe related models and motivate the calculations in this work.

The simplest continuum field theory in the $O(3)$ universality class is the theory of a scalar field $\vec \phi$ transforming in the fundamental representation of $O(3)$ with Lagrangian
\be
\label{eq:o3theory}
\cL &= \frac 1 2 |\ptl \vec \phi|^2 + \frac {1}{2} m^2 |\vec \phi|^2 + \frac{g}{4!} |\vec\phi|^4 \,.
\ee
A large negative mass-squared for the scalar induces spontaneous symmetry breaking and leads to the ordered phase, while a large positive mass-squared leads to the disordered phase. The critical point is achieved by tuning the UV mass so that the infrared (IR) correlation length diverges. The $\beta$ function of the coupling $g$ has been computed in the $\varepsilon-$expansion and in a fixed-dimension scheme. After a Borel-resummation, both methods predict the existence of an IR stable fixed point. We will review these results in the next sections.

The IR limit of the above field theory captures the same physics as the Heisenberg model. This model consists of a lattice of classical spins $\vec S_i$, which can take values on a three-dimensional sphere. The Hamiltonian has only nearest-neighbor interactions:
\be
\label{eq:Heisenbergmodel}
\mathcal H = - J \sum_{\langle i,j\rangle}  \vec S_i\cdot \vec S_j + H \sum_i S_i^3\,,
\ee
where we also introduced an external magnetic field $H$ in the third ($z$)-direction.
When the parameter $J$ is positive, the ground state corresponds to all spins aligned, corresponding to ferromagnets. When $J<0$, the energy is minimized when neighboring spins are anti-aligned, corresponding to antiferromagnets.
 
For small $J$, the line $H=0$ separates a ferromagnetic phase from the paramagnetic one. This line represents a first-order transition and terminates at a value $J=J_c$, where the correlation length of the system diverges, and the transition becomes second order. For $J>J_c$, there is only a disordered phase. At $J=J_c$, the theory in the IR is in the same universality class of the field theory defined in (\ref{eq:o3theory}). The critical exponents are related to operator dimensions at the fixed point as
\be
\Delta_\phi = \frac{1+\eta}2 \,,\qquad \Delta_s = 3-\frac1{\nu}\,,\qquad \Delta_t = 3-Y_2 \,.
\ee 
Here, $s\sim |\vec{\phi}|^2$ denotes the lowest-dimension singlet scalar, while $t_{ij} \sim( \phi_i\phi_j -\text{trace})$ denotes the lowest rank-2 scalar. More generically the exponents $Y_r$ are associated to the dimensions of the lowest rank-$r$ scalar operator.\footnote{Sometimes in the literature they are replaced by the crossover exponents $\phi_r =\nu Y_r $. }
In the \ON model, the dimension of the lowest traceless symmetric operator $t$ describes the instability of the theory against anisotropic perturbations. Because of this, it plays an important role in the description of multicritical phenomena. For instance, the critical behavior near a bicritical point where two critical lines with $\Ononewithout$ and $\Ontwowithout$ symmetry meet givies rise to a critical theory with enlarged $\Ononetwowithout$ symmetry. 

The Hamiltonian (\ref{eq:Heisenbergmodel}) is a simplified model of magnetic interactions, since in a real crystalline solid other interactions are present. For instance, the crystal lattice structure could give rise to magnetic anisotropy. In cubic-symmetric lattices this effect produces an  interaction localized at each lattice point $i$ of the form $\sum_{k=1}^3 (S_i^k)^4$. This perturbation breaks the $O(3)$ global symmetry of the Heisenberg Hamiltonian, and therefore it cannot be generated by an RG transformation. As such, the IR fixed point of (\ref{eq:Heisenbergmodel}) will be described by an $O(3)$ invariant CFT.

\subsubsection{\ON vs.\ multi-critical models}
\label{sec:ONvscubic}

The $O(3)$ model described above can be generalized to \ON by promoting $\vec{\phi}$ to an $N$ component field. 
We can also consider the closely related cubic model, which describes the continuum limit of the Hamiltonian (\ref{eq:Heisenbergmodel}) with the addition of the \ON breaking term $\sum_{k=1}^N (S_i^k)^4$. This interaction is indeed invariant under the symmetries of a hyper-cubic lattice, namely permutations and reflection of the three axes. The field $\phi_i$, $i =1,\ldots, N$, transforms in the fundamental representation of the permutation group $\mathcal S_N$. Moreover, each component is odd under a reflection of the corresponding axis. The composition of these transformations gives rise to the hypercubic symmetry group $C_N = \mathds Z_N \rtimes \mathcal S_N$.

Compared to (\ref{eq:o3theory}), the Lagrangian of the hypercubic model has an additional term in the potential:
\be
\label{eq:cubictheory}
\cL &= \frac 1 2 \sum_{i=1}^N \left((\ptl_\mu \phi_i)^2 +  m^2\phi_i^2 \right)+ \frac{g}{4!} \left(\sum_{i=1}^N \phi_i^2\right)^2  +  \frac{h}{4!} \sum_{i=1}^N \phi_i^4\,.
\ee
The computation of the two $\beta$-functions $\beta_g$ and $\beta_h$ reveals the existence of four fixed points \cite{PhysRevB.8.3323}: the trivial fixed point ($g=h=0$), the $N$ decoupled copies of the Ising model ($h\neq0$, $g=0$), the \ON fixed point ($g\neq0$, $h=0$) and the cubic model ($g\neq0$, $h\neq0$). It is straightforward to see that the first two are unstable since the quartic operator parametrized by $g$ is relevant in both theories.\footnote{The case of $N$ decoupled Ising model corresponds to products of operators $\epsilon_i\sim \phi_i^2$ belonging to different copies where $2\Delta_\epsilon<3$ in the Ising model.}
Determining which one of the other two fixed points is stable is a more complicated issue, and it turns out to be $N$ dependent. 

One way to rephrase the above question is to notice that the additional term in (\ref{eq:cubictheory}) can be rewritten as
\be
\sum_{i=1}^N \phi_i^4 = \sum_{i=1}^N t_4^{iiii} + \frac{3}{N+2} \left(\sum_{i=1}^N \phi_i^2\right)^2\,,
\ee
where $t_4^{ijkl}$ is the traceless symmetric combination of four fields. The added term in the potential, in \ON notation, can be written as a combination of a rank-4 field and a singlet. We know that the singlet is irrelevant at the \ON fixed point, by definition. Thus the stability of the \ON fixed point or the cubic point is linked to the value of the dimension of the operator $t_4$.

In the ${O}(2)$ model the operator $t_4$ is irrelevant. A simple proof of this is to notice that for $N=2$, as long as $h\neq 0$, the cubic Lagrangian can be mapped in the Lagrangian of two decoupled Ising models. This cubic fixed point coincides with the decoupled Ising fixed point, which is unstable. Field theory and Monte Carlo determinations of the dimension of $t_4$ agree with this argument. This is also consistent with the assumptions made in \cite{Chester:2019ifh}. \\
On the contrary, at large $N$, the operator $t_4$ is relevant, and the cubic fixed point is stable. Thus, it is important to know at which value $N=N_c>2$ the operator $t_4$ becomes relevant. 

A second closely related model is the multi-critical point with $\Ononetwodirectwithout$ symmetry \cite{PhysRevLett.33.813}. A field theory description is given in terms of two sets of scalar fields $\vec \phi_1$ and $\vec \phi_2$, transforming respectively in the fundamental representation of $\Ononewithout$ and $\Ontwowithout$, with Lagrangian:
 \be
\label{eq:multicriticaltheory}
\cL &= \frac 1 2 \sum_{i=1}^2 |\ptl_\mu \vec \phi_i|^2 +  \frac{g_1}{4!} \left(|\vec\phi_1|^2\right)^2 +  \frac{g_2}{4!}   \left(|\vec\phi_2|^2\right)^2 +   \frac{h}{4} |\vec\phi_1|^2 |\vec\phi_2|^2\,,
 \ee
where we have already set to zero all the mass terms. The analysis of the perturbative $\beta$ functions shows the existence of six fixed points. Some we already know: the free one ($g_i=0$, $h=0$), the two Wilson Fisher fixed points ($g_1\neq0$, $g_2=h=0$ and same with $1\leftrightarrow2$), the decoupled fixed point (DFP, $g_i\neq0$, $h=0$), the symmetry enhanced $\Ononetwowithout$ Wilson Fisher fixed point, and lastly the biconal fixed point (BFP). The latter one also has all couplings nonvanishing, but the global symmetry is not enhanced.

The problem of understanding the stable fixed point can be again reduced to studying the (ir)relevance of given deformations in the various CFTs. For instance, by inspecting the dimension of the composite operator built out of the lowest dimension scalar singlets in $\Ononewithout$ and $\Ontwowithout$ theories, one can conclude that the DFP is stable for any $N=n_1+n_2\geq 4$. It is unstable for $N\leq 3$, although the perturbation is close to being marginal.\footnote{The most precise bootstrap determination \cite{Kos:2016ysd,Chester:2019ifh} gives 
$ \Delta_{[s_{\mathds Z_2} s_{\Otwowithout}]} = \Delta_{s_{\mathds Z_2}} + \Delta_{s_{\Otwowithout}} =
2.92398(23) <3$.}

The issue of stability of \ON vs.~the BFP is again related to the dimension of a certain operator. In the Lagrangian formulation, this is a combination of quartic interactions. At the \ON fixed point this term is mapped in a combination of the second-lowest rank-0 ($S'$), second-lowest rank-2 ($t_2'$) and leading rank-4 scalar operator $t_4$. If any of these operators is relevant, then the \ON fixed point is unstable. While the former two are known to be always irrelevant for any $N$, the latter is the object of investigations. In particular, if $\Delta_{t_4}<3$ for $N=3$, then among the fixed points, the BFP will be the stable one.

\subsubsection{Field theory results}

\begin{table*}[ht]
\centering
\begin{tabular}{@{}cl|cc@{}}
	\toprule
CFT data & method & value & ref \\
	\midrule
$\Delta_s$ & $d=3$ exp & 1.5840(14) &\cite{Jasch_2001} \\ 
 & $\varepsilon$-exp & 1.580(11)  & \cite{Guida_1998}\\
 & HT & 1.603(4)  & \cite{Butera_1998}\\
	 \midrule
 $\Delta_\phi$ & $d=3$ exp & 0.5175(4) & \cite{Jasch_2001} \\
 & $\varepsilon$-exp &0.5188(23) & \cite{Guida_1998} \\
 & HT &0.5180(35) & \cite{Butera_1998} \\
	 \midrule
 $\Delta_t$ & $d=3$ exp & 1.20(3)  & \cite{Calabrese_2003} \\
 & $\varepsilon$-exp & 1.210(3) & \cite{Calabrese_2003}\\
 & HT & 1.24(2) & \cite{PhysRevB.10.2088}\\
 \midrule
$\Delta_{t_4}$ & $d=3$ exp & 2.987(6) & \cite{Carmona:1999rm} \\
 & $\varepsilon$-exp & 2.997(4) & \cite{Carmona:1999rm}\\
\bottomrule 
\end{tabular}
	\caption{Comparison of field theory results using various techniques: fixed-dimensional expansion in three dimensions ($d=3$ exp), epsilon expansion ($\varepsilon$-exp) and high temperature expansion (HT). We denote the leading rank-0, rank-1, rank-2, and rank-4 scalars by $s,\f,t,t_4$, respectively. 
	Another estimate of $\Delta_t$ in the fixed-dimensional expansion can be found in \cite{Calabrese_2002} in terms of the crossover exponents $\phi_T = Y_2 \nu$, with $Y_2 = 3-\Delta_t$. We do not report it here because the errors depend on the value of $\nu$ used.
	\label{tab:FTresults}}
\end{table*}

Both the $O(3)$ model and the cubic model have been extensively studied using different expansion techniques. $\beta$-functions for these models are known up to high order in both the $\epsilon$-expansion and fixed-dimension expansion, and critical exponents have been computed by Borel resumming the respective series.\footnote{Both approaches are based on a perturbative expansion in the quartic interaction $g$ up to a certain loop order. In the fixed dimension approach one works directly in $d=3$ dimension and looks for solutions of the Borel resummed $\beta$-function $\beta^\text{BR}(g_*)=0$. Critical exponents are then computed as $\nu^\text{BR}(g_*)$. In order to remove the divergences one imposes suitable renormalization conditions. Historically the term ``fixed dimension" refers to renormalization conditions at zero momentum; the use of a minimal subtraction scheme is instead called a ``minimal subtraction scheme without $\varepsilon$-expansion". In the proper $\varepsilon$-expansion approach one works in $d=4-\varepsilon$ dimensions and solves the condition $\beta(g_*)=0$ order by order in $\varepsilon$. Plugging the solution $g_*(\varepsilon)$ into the expression for the critical exponent one gets a series in $\varepsilon$ that can be Borel resummed. The final critical exponents are then computed as $\nu^{\varepsilon\text{-BR}}(\varepsilon=1)$.} We report in table~\ref{tab:FTresults} the latest results obtained with field theory techniques.

The question of stability of fixed points has also been discussed in the literature. As we discussed in the previous section, this question can be addressed in two ways: by computing the dimension of the lowest dimension rank-4 scalar in the $O(3)$ model, or by computing the value $N_c$ at which the dimension of the second-lowest rank-0 scalar in the cubic model becomes exactly marginal.\footnote{The lowest dimension one corresponds to the mass deformation and is always relevant; the second-lowest corresponds to a combination of the two quartic interactions. The orthogonal combination is related to the Lagrangian operator via the equation of motion and is irrelevant.}  Results from both methods support the conclusion that $O(3)$ is unstable while the cubic model is stable. The formula for $N_c$ in the $\varepsilon$-expansion is \cite{Kleinert_1995,Varnashev:1999ze}:
\begin{equation}
N_c = 4 - 2\varepsilon + 2.58847559\varepsilon^2 - 5.87431189\varepsilon^3 + 16.82703902 \varepsilon^4 + O[\varepsilon^5]\,,
\end{equation}
and after resummation gives $N_c = 2.89(2)$.

Analysis of the $\varepsilon$-expansion or fixed-dimension perturbative series in the cubic model \cite{Folk_2000,Calabrese_2003,Manuel_Carmona_2000} shows that the critical exponents of the two models are very close:\footnote{Field theory estimates have also been obtained in \cite{Tissier:2002zz,Adzhemyan:2019gvv}}
\begin{equation}
\label{eq:exponentsdifference}
\nu_{\Othreewithout} -\nu_\text{cubic}  = 0.0003(3)\,,\qquad   \eta_{\Othreewithout}- \eta_\text{cubic}  = 0.0001(1)\,.
\end{equation}
These differences are much smaller than the typical experimental error (e.g., \cite{Pelissetto:2000ek}). This makes distinguishing the two models experimentally very challenging. Curiously, the first few terms in the of the $\varepsilon$-expansion of the critical exponents in \eqref{eq:exponentsdifference} are quite different, and it is only after the Borel resummation that the two values appear quite close.

Similarly, also the biconal $\mathds Z_2 \times \Otwowithout$ model and the \Othree critical exponents are very close, as the flow connecting the two is driven by the same almost marginal operator as in the cubic case:
\begin{equation}
\label{eq:exponentsdifferenceBFP}
|\nu_{\Othreewithout} -\nu_\text{BFP} | \lesssim 0.001\,,\qquad   |\eta_{\Othreewithout}- \eta_\text{BFP}|  = 0.0005 \,,\qquad  |\eta_{\Othreewithout}- \eta'_\text{BFP}|  = 0.0001\,, 
\end{equation}
where $\eta_\text{BFP}$ and $\eta'_\text{BFP}$ correspond to the two relevant order parameters charged respectively under $\mathds Z_2$ or $\Otwowithout$.

\subsubsection{Monte Carlo results}

Using Monte Carlo (MC) techniques, it is possible to obtain precise estimates of the critical exponents for both the \Othree model and the cubic model, as well as information about their stability. Such determinations can also be improved when combined with finite-size scaling (FSS) or high-temperature expansion (HT) methods. A precise determination of the $\nu$ and $\eta$ critical exponents was made using MC and FSS methods in \cite{Hasenbusch_2001}. A more precise analysis combining MC with HT techniques was carried out in \cite{Campostrini:2002ky}, while a more precise MC and FSS study was performed in \cite{Hasenbusch_2011}. A very precise MC and FSS analysis of an icosahedral model, as well as improved MC and HT analyses were recently presented in \cite{Hasenbusch:2020pwj}. Several other less precise determinations can be found in \cite{Pelissetto:2000ek}. The dimensions relevant to anisotropic perturbations of rank-2,3,4 were computed in \cite{Hasenbusch_2011, Caselle:1997gf} using MC and FSS methods, and support the conjecture that the \Othree model is unstable under cubic deformations.  These results are summarized in table~\ref{tab:results}.

\subsubsection{The conformal bootstrap}

Three dimensional \Othree models have been studied with bootstrap methods in a series of papers \cite{Kos:2013tga,Kos:2015mba,Kos:2016ysd}, first by considering the correlation function $\<\phi_i\phi_j \phi_k\phi_l\>$, where $\phi_i$ is the lowest-dimension scalar transforming in the vector representation of \ON, and then by also including correlation functions involving the lowest-dimension singlet scalar $s$. The most precise determination of the critical exponents was obtained in \cite{Kos:2016ysd}, which isolated a three dimensional region in the space $\{\Delta_\phi,\Delta_s,\lambda_{sss}/\lambda_{\phi\phi s}\} = \{0.51928(62), 1.5957(55), 1.205(9)\}$, under the assumption that $\phi_i$ and $s$ are the unique relevant scalar operators in their representations. In addition, by scanning over this island, \cite{Kos:2016ysd} determined the magnitude of the leading OPE coefficient to be $\lambda_{\f\f s} = 0.5244(11)$. 
  
Theories invariant under the cubic symmetry group were also studied using bootstrap methods using single correlators  \cite{Rong:2017cow, Stergiou:2018gjj}, and mixed correlators \cite{Kousvos:2018rhl, Kousvos:2019hgc}. In particular \cite{Stergiou:2018gjj} analyzed the bootstrap equations assuming the hypercubic symmetry group $C_N = \mathds Z_N \rtimes \mathcal S_N$ and observed a series of kinks for various values of $N$. However, the locations of the kinks in the singlet sector were degenerate with the \ON kinks (and hence compatible with (\ref{eq:exponentsdifference})), likely reflecting a symmetry enhancement in the extremal bootstrap solutions~\cite{Poland:2011ey,Nakayama:2017vdd,Li:2018lyb,Li:2020bnb,Li:2020tsm}, while the bounds in other sectors did not seem to be saturated by the cubic model. The mixed-correlator analysis of \cite{Kousvos:2018rhl,Kousvos:2019hgc} also did not manage to isolate the cubic model but rather found evidence of a new theory, called the ``Platonic CFT," with cubic symmetry and operator dimensions not matching any known CFT.

In this work, we study the \Othree model with numerical bootstrap techniques using a larger system of correlation functions than before: in addition to $\phi_i$ and $s$, we incorporate the lowest-dimension rank-2 scalar $t_{ij} \sim \phi_{(i}\phi_{j)}$. This setup is similar to the one leading to the successful results obtained in \cite{Chester:2019ifh} for the \Otwo model. Following the strategy detailed in \cite{Chester:2019ifh}, we first scan over the three operator dimensions $\{\De_\f,\De_s,\De_t\}$ and the OPE coefficients $\{\l_{sss}, \l_{\f\f s}, \l_{tts}, \l_{\f\f t}, \l_{ttt}\}$ (or more precisely their ratios) and we determine a three dimensional island in the space of operator dimensions, along with an associated allowed set of OPE coefficient ratios. Next, we compute upper and lower bounds on the magnitude $\lambda_{\f\f s}$, as well as on the current and stress-tensor central charges $C_J$ and $C_T$. Finally, we enlarge the parameter space to include one more parameter: the dimension of the lowest rank-4 scalar $\Delta_{t_4}$. Using the new \texttt{tiptop} algorithm, which we describe in section \ref{sec:tiptop}, we carve out the allowed region in the enlarged four-dimensional space and obtain an upper bound on $\Delta_{t_4}$.

\subsection{Structure of this work}

The remainder of this work is structured as follows. In section~\ref{sec:o3mod} we describe the crossing equations and relevant \Othree representation theory. In section~\ref{sec:tiptop} we describe the new \texttt{tiptop} algorithm that we use in order to bound $\Delta_{t_4}$. In section~\ref{sec:results} we describe the results of our numerical bootstrap calculations and in section~\ref{sec:conclusions} we describe directions for future research. Various appendices describe the code availability, software setup, details about our tensor structures, and give a list of allowed and disallowed points that we have computed.

\section{The $O(3)$ model}
\label{sec:o3mod}

\subsection{Crossing equations}
\label{crosssec}

We begin by describing the representation theory of $\Othreewithout =  \mathds{Z}_2 \times \SOthreewithout $. We label the irreducible representations ${\bf q}^{\pm}$ of \Othree by the usual \SOthree rank $\bf q$ tensor of dimension $2q+1$  for $q\in \frac12\mathds{Z}_{\geq0}$, as well as the $\mathds{Z}_2$ parity $\pm$. Tensor products of these irreps are given by
\es{tenprod}{
{\bf q_1}^{\pm}\otimes {\bf q_2}^{\pm} = \bigoplus_{q_a=|q_1-q_2|}^{q_1+q_2}{\bf q_a}^+\,,\qquad \qquad{\bf q_1}^{\pm}\otimes {\bf q_2}^{\mp} = \bigoplus_{q_a=|q_1-q_2|}^{q_1+q_2}{\bf q_a}^-\,,
}
where if ${\bf q_1}^{\pm}={\bf q_2}^{\pm}$, then the even/odd ${\bf q_a}$ are in the symmetric/antisymmetric part of the tensor product.

Operators $\cO_{\bf q^\pm}(x)$ in the irrep ${\bf q^\pm}$ can be written in terms of \SOthree fundamental indices $i=1,2,3$ as rank-$q$ symmetric traceless tensors $\cO_\pm^{i_1\dots i_q}(x)$ with the extra $\mathds{Z}_2$ labels $\pm$. Four-point functions of scalar operators $\varphi_{\pm}^{i_1\dots i_{q}}(x)$ can be expanded in the $s$-channel in terms of conformal blocks as\footnote{Our conformal blocks are normalized as in the second line of table 1 in \cite{Poland:2018epd}.}
 \be
 \label{4point}
&\left\langle  \varphi_{\pm_1}^{i_1\dots i_{q_1}}(x_1)  \varphi_{\pm_2}^{j_1\dots j_{q_2}}(x_2)   \varphi_{\pm_3}^{k_1\dots k_{q_3}}(x_3)   \varphi_{\pm_4}^{l_1\dots l_{q_4}}(x_4)  \right\rangle   \nn\\
&=\frac{\left(\frac{x_{24}}{x_{14}}\right)^{{\Delta_{12}}}  \left(\frac{x_{14}}{x_{13}}\right)^{{\Delta_{34}}} }{x_{12}^{\Delta_1+\Delta_2}x_{34}^{\Delta_3+\Delta_4}} \sum_{\cO}(-1)^\ell\lambda_{\varphi_1\varphi_2\cO}\lambda_{\varphi_3\varphi_4\cO}T^{\cR,{ i_1\dots i_{q_1} ,j_1\dots j_{q_2} ,k_1\dots k_{q_3}, l_1\dots l_{q_4}}}_{\cR_1\cR_2\cR_3\cR_4}g^{\Delta_{12},\Delta_{34}}_{\Delta,\ell}(u,v) \,,
\ee
 where $\Delta_{ij}\equiv\Delta_i-\Delta_j$, the conformal cross ratios $u,v$ are 
 \es{uv}{
 u \equiv \frac{{x_{12}^2x_{34}^2}}{{x_{13}^2x_{24}^2}},\qquad v \equiv \frac{{x_{14}^2x_{23}^2}}{{x_{13}^2x_{24}^2}}\,,
 }
and the operators $\cO$ that appear both OPEs $\varphi_1\times\varphi_2$ and $\varphi_3\times\varphi_4$ have scaling dimension $\Delta$, spin $\ell$, and transform in an irrep $\cR$ that appears in both the tensor products $\mathcal{R}_1\otimes\mathcal{R}_2$ and $\mathcal{R}_3\otimes\mathcal{R}_4$. For each $\cR$, the \SOthree structure $T^\cR$ can be constructed using the \SOthree Casimir and normalized to give consistent OPE coefficients under crossing using the free theory as described in appendix \ref{tensApp}. The $\mathds{Z}_2$ irrep of $\cO$ follows from trivial multiplication of $\pm_1$ and $\pm_2$, and so does not require a structure. If $\varphi_1=\varphi_2$ (or $\varphi_3=\varphi_4$), then Bose symmetry requires that $\cO$ have only even/odd $\ell$ for $\mathcal{R}$ in the symmetric/antisymmetric product of $\mathcal{R}_1\otimes\mathcal{R}_2$ (or $\mathcal{R}_3\otimes\mathcal{R}_4$).

We are interested in four-point functions of the lowest dimension scalar operators transforming in the ${\bf0^+}$, ${\bf1^-}$, and ${\bf2^+}$ representations, which we will denote following \cite{Kos:2013tga,Kos:2015mba,Kos:2016ysd} as $s$, $\phi$, and $t$, respectively.\footnote{The singlet $S$, traceless symmetric $T$, vector $V$, and antisymmetric $A$ irreps considered in previous \ON bootstrap papers \cite{Kos:2013tga,Kos:2015mba,Kos:2016ysd} correspond for \Othree to the ${\bf 0^+}$, ${\bf 2^+}$, ${\bf 1^-}$, and ${\bf 1^-}$ irreps, respectively, where now $A\cong V$.} These operators are normalized via their two point functions as
\es{2points}{
\langle s(x_1) s(x_2) \rangle=\frac{1}{x_{12}^{2\Delta_s}}\,,\quad
\langle \phi^{i}(x_1) \phi^j(x_2) \rangle&=\frac{\delta^{ij}}{x_{12}^{2\Delta_\phi}}\,,\quad
\langle t^{i_1i_2}(x_1) t^{j_1j_2}(x_2) \rangle=\frac{\delta^{i_1j_1}\delta^{i_2j_2}}{x_{12}^{2\Delta_t}}\,,\\
}
where $x_{12}\equiv |x_1-x_2|$ and all indices of the same letter should be symmetrized with their trace removed. In table \ref{table} we list the 4-point functions of $s$, $\phi$, and $t$ that are allowed by \Othree symmetry\footnote{If we had just \SOthree symmetry, then in addition to these 4-point functions we would also have $\langle s\phi\phi\phi\rangle$, $\langle stt\phi\rangle$, $\langle t\phi\phi\phi\rangle$, and $\langle \phi ttt\rangle$, which can be constructed using the \SOthree invariant tensor $\epsilon_{ijk}$. These correlators give an additional 9 crossing equations for 37 total. As discussed in \cite{Kos:2015mba}, to distinguish between \SOthree and $O(3)$, one needs to set some of the OPE coefficients in these additional correlators to be nonzero. Otherwise, the extra crossing equations have no effect.} and whose $s$ and $t$-channel configurations lead to independent crossing equations, along with the irreps and spins of the operators that appear in the OPE, and the number of crossing equations that they yield.
\begin{table}
\begin{center}
\begin{tabular}{@{}c|c|c|c@{}}
	\toprule
 Correlator& $s$-channel & $t$-channel&Eqs\\
 \midrule 
$\langle\phi\phi\phi\phi\rangle$&  $(\ell^+,{\bf0^+})$, $(\ell^-,{\bf1^+})$, $(\ell^+,{\bf2^+})$&same&  3   \\
 \midrule
 $\langle tttt\rangle$&   $(\ell^+,{\bf0^+})$, $(\ell^-,{\bf1^+})$, $(\ell^+,{\bf2^+})$, $(\ell^-,{\bf3^+})$, $(\ell^+,{\bf4^+})$&same&  5   \\
 \midrule
 $\langle t\phi t\phi\rangle$&   $(\ell^\pm,{\bf1^-})$,$(\ell^\pm,{\bf2^-})$, $(\ell^\pm,{\bf3^-})$ &same&  3   \\
 \midrule
 $\langle tt\phi\phi\rangle$&   $(\ell^+,{\bf0^+})$, $(\ell^-,{\bf1^+})$&$(\ell^\pm,{\bf1^-})$,$(\ell^\pm,{\bf2^-})$, $(\ell^\pm,{\bf3^-})$&  6   \\
 \midrule
 $\langle ssss\rangle$&  $(\ell^+,{\bf0^+})$&same&  1   \\
 \midrule
 $\langle\phi s\phi s\rangle$&  $(\ell^\pm,{\bf1^-})$&same&  1   \\
 \midrule
 $\langle tsts\rangle$&  $(\ell^\pm,{\bf2^+})$&same&  1   \\
 \midrule
 $\langle ttss\rangle$&  $(\ell^+,{\bf0^+})$& $(\ell^\pm,{\bf2^+})$ &  2   \\
 \midrule
 $\langle\phi\phi ss\rangle$&   $(\ell^+,{\bf0^+})$& $(\ell^\pm,{\bf1^-})$ &  2   \\
 \midrule
 $\langle\phi s\phi t\rangle$&   $(\ell^\pm,{\bf1^-})$& same &  1   \\
 \midrule
 $\langle \phi\phi st\rangle$&   $(\ell^+,{\bf2^+})$ & $(\ell^\pm,{\bf1^-})$   &  2   \\
 \midrule
 $\langle stt t\rangle$&   $(\ell^\pm,{\bf2^+})$& same &  1   \\
 \bottomrule
\end{tabular}
\caption{Four-point function configurations that give independent crossing equations under equating their $s$- and $t$-channels, along with whether even or odd spins $\ell^{\pm}$ appear for each irrep in each channel, and the number of crossing equations that each configuration yields.}
\label{table}
\end{center}
\end{table}
 These 4-point functions can be written explicitly as in \eqref{4point}, where the explicit \SOthree structures  $T^\cR$ are computed in appendix \ref{tensApp}. Equating each of these $s$-channel 4-point functions with their respective $t$-channels yields the crossing equations
  \es{crossing3}{
&0=\sum_{ \cO_{\bf0^+},\ell^+}   \begin{pmatrix} \lambda_{ss\cO_{\bf0^+}} & \lambda_{\phi\phi\cO_{\bf0^+}} & \lambda_{tt\cO_{\bf0^+}} \end{pmatrix} \vec V_{{\bf0^+},\Delta,\ell^+}  \begin{pmatrix}  \lambda_{ss\cO_{\bf0^+}} \\ \lambda_{\phi\phi\cO_{\bf0^+}} \\ \lambda_{tt\cO_{\bf0^+}}  \end{pmatrix} \\
& +
 \sum_{  \cO_{\bf1^+},\ell^-}  \begin{pmatrix} \lambda_{ \phi\phi\cO_{\bf1^+}} &  \lambda_{tt\cO_{\bf1^+}}\end{pmatrix} \vec V_{{\bf 1^+},\Delta,\ell^-}  \begin{pmatrix}  \lambda_{ \phi\phi\cO_{\bf1^+}} \\  \lambda_{tt\cO_{\bf1^+}}\end{pmatrix} + \sum_{ \cO_{\bf1^-},\ell^\pm}  \begin{pmatrix} \lambda_{ t\phi \cO_{\bf1^-}} & \lambda_{\phi s \cO_{\bf1^-}} \end{pmatrix} \vec V_{{\bf 1^-},\Delta,\ell^\pm}  \begin{pmatrix} \lambda_{ t\phi \cO_{\bf1^-}} \\ \lambda_{ \phi s \cO_{\bf1^-}} \end{pmatrix}
  \\
 &+   \sum_{ \cO_{\bf2^+},\ell^+}  \begin{pmatrix} \lambda_{\phi\phi \cO_{\bf2^+}} &\lambda_{tt \cO_{\bf2^+}}  &\lambda_{ts \cO_{\bf2^+}}  \end{pmatrix} \vec V_{{\bf 2^+},\Delta,\ell^+}  \begin{pmatrix}  \lambda_{\phi\phi \cO_{\bf2^+}} \\ \lambda_{tt \cO_{\bf2^+}}  \\ \lambda_{ts \cO_{\bf2^+}}  \end{pmatrix}  +   \sum_{  \cO_{\bf2^+},\ell^-} \lambda_{ t s\cO_{\bf2^+}}^2 \vec V_{{\bf 2^+},\Delta,\ell^-}  \\
&  +   \sum_{  \cO_{\bf2^-},\ell^\pm}  \lambda_{ t \phi \cO_{\bf2^-}}^2  \vec V_{{\bf 2^-},\Delta,\ell^\pm}   +   \sum_{ \cO_{\bf3^+},\ell^-}  \lambda_{tt \cO_{\bf3^+}}^2  \vec V_{{\bf 3^+},\Delta,\ell^-} +   \sum_{ \cO_{\bf3^-},\ell^\pm}  \lambda_{t \phi \cO_{\bf3^-}}^2  \vec V_{{\bf 3^-},\Delta,\ell^\pm} + \sum_{ \cO_{\bf4^+},\ell^+}  \lambda_{ tt\cO_{\bf4^+}}^2  \vec V_{{\bf 4^+},\Delta,\ell^+}\,,
 }
 where $\ell^\pm$ denotes which spins appear, and the $V$'s are 28-dimensional vectors of matrix or scalar crossing equations that are ordered as table \ref{table} and written in terms of
\es{Fdefine}{
F^{ij,kl}_{\mp,\Delta,\ell}(u,v)=v^{\frac{\Delta_k+\Delta_j}{2}}g_{\Delta,\ell}^{\Delta_{ij},\Delta_{kl}}(u,v)\mp u^{\frac{\Delta_k+\Delta_j}{2}}g_{\Delta,\ell}^{\Delta_{ij},\Delta_{kl}}(v,u)\,.
}
The explicit form of the $V$'s is given in the attached Mathematica notebook.\footnote{These crossing equations can also be derived using the software package {\tt autoboot} \cite{Go:2019lke}.}

\subsection{Ward identities}
\label{assumpSec}

The OPE coefficients of $J^\mu$ and $T^{\mu\nu}$ are constrained by Ward identities in terms of the two-point coefficients $C_J$ and $C_T$. In our conventions, we have
\be
\l_{\cO \cO T}^2 = \frac{\De_\cO^2}{3C_T/C_T^\mathrm{free}}\,,\quad \l_{\cO\cO J}^2 = \frac{q_\cO^2}{2C_J/C_J^\mathrm{free}}\,,
\ee
where $C_{J,T}^\mathrm{free}$ are the two-point coefficients of $J$ and $T$ in the free \Othree model described in appendix \ref{tensApp}. Thus, the contribution of these operators to the crossing equation can be parametrized purely in terms of $C_T$ and $C_J$, together with the dimensions and charges of the external scalars $\f,s,t$.

\section{The \texttt{tiptop} algorithm}
\label{sec:tiptop}

While our primary search for the \Othree bootstrap island will follow the 
same methods and software tools used for the \Otwo model described in~\cite{Chester:2019ifh}, we will
also need to compute the maximum value of the scaling dimension $\Delta_{t_4}$ over this island. 
This employs a new search strategy and software implementation that we describe in this section.

\subsection{Software and algorithm}
\texttt{tiptop} is a program to assist in finding the maximum value of
a coordinate achieved in a region in $(N+1)$-dimensional space, where
testing whether a point is in the region is computationally expensive.
Given a set of inside (allowed), outside (disallowed), and unknown
points, \texttt{tiptop} generates successive points to narrow down the
boundary of the top of the region.  It is meant to be invoked by a
driver that takes these points, computes whether they are allowed, and
then asks for more points to check.  \texttt{tiptop} is freely
available (see Appendix \ref{tiptop:availability}).

The number of dimensions is arbitrary but fixed at compile time.  For
concreteness and ease of visualization, we assume that $N+1=3$ for the
rest of this discussion, where the dimensions are
$\Delta_{\phi}, \Delta_{s},$ and $\Delta_{\text{gap}}$.  The algorithm
operates unchanged for higher dimensions.

We start with at least one allowed point, a cloud of disallowed
points, a cloud of points that are in-progress, and a maximum gap
($\Delta_{\text{max\_gap}}$).  In-progress points are points that the
driver already knows about and is working on, but does not yet know if
they are allowed.  For example, those calculations may have been
submitted as calculations to an HPC cluster but not yet completed.

We assume that there are no allowed points with
$\Delta_{\text{gap}} \geq \Delta_{\text{max\_gap}}$.  We also assume
that islands only shrink at larger gaps.  This allows us to assert
that if a point is disallowed at one gap, it will continue to be
disallowed at larger gaps.

The last assumption is that each $N$-dimensional island at a fixed
value of $\Delta_{\text{gap}}$ is convex and simply connected, so each
island never becomes a horseshoe or splits into two pieces.  We have
observed this behavior for a wide variety of theories, as long as the
theory is well approximated.

The basic outline of the algorithm for generating points is:
\begin{itemize}
\item Set $\Delta_{\text{allowed}}$ to the largest
  $\Delta_{\text{gap}}$ with an allowed point.
\item Explore parameters at $\Delta_{\text{allowed}}$ to find the
  size of the island there.  If there are any corners of parameter
  space left to map out, return one point from there (Section
  \ref{subsec:exploring}).
\item If the island at $\Delta_{\text{allowed}}$ is thoroughly
  mapped out, generate one point at a higher gap (Section \ref{subsec:jumping}).
\end{itemize}

The non-gap dimensions ($\Delta_{\phi}, \Delta_{s}$)
are represented as regular floating-point numbers, while the gap dimension
($\Delta_{\text{gap}}$) has been rescaled to a 64 bit
integer.  This reduces numerical errors where two points at very
similar gaps are mistakenly considered to be at the same gap.

\texttt{tiptop} will not return a point in two cases:
\begin{itemize}
\item The current gap $\Delta_{\text{allowed}}$ might be fully
  explored, but it needs to know the outcome of some in-progress
  points to be sure.
  
\item There are no valid larger gaps left.  For example, consider the case
  where $\Delta_{\text{allowed}}=10000$ and \texttt{tiptop} has ruled
  out any jumps to $\Delta_{\text{gap}}=10001$.  There are no integers
  between 10000 and 10001, so the algorithm terminates.
\end{itemize}

\subsection{Exploring the current gap}
\label{subsec:exploring}

\subsubsection{Rescaling}
\label{subsubsec:Rescaling}

\begin{figure}

\begin{minipage}[t]{0.49\textwidth}%
\includegraphics[width=1\textwidth]{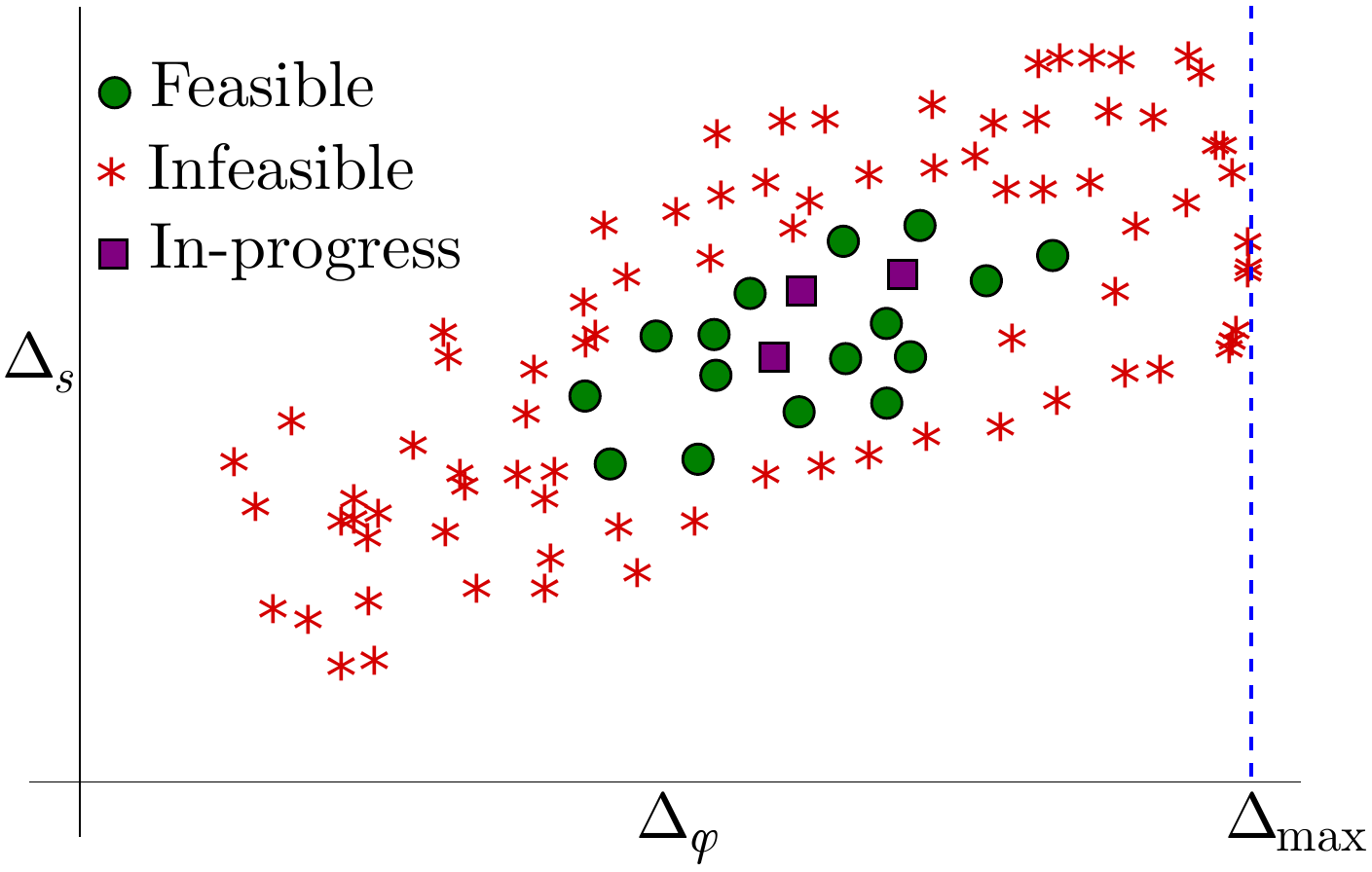}
\caption{\label{tiptop:box} The max coordinates $\Delta_{\max}$ for a collection of allowed, disallowed, and in-progress points.}
\end{minipage}\hfill{}%
\begin{minipage}[t]{0.49\textwidth}%
\includegraphics[width=1\textwidth]{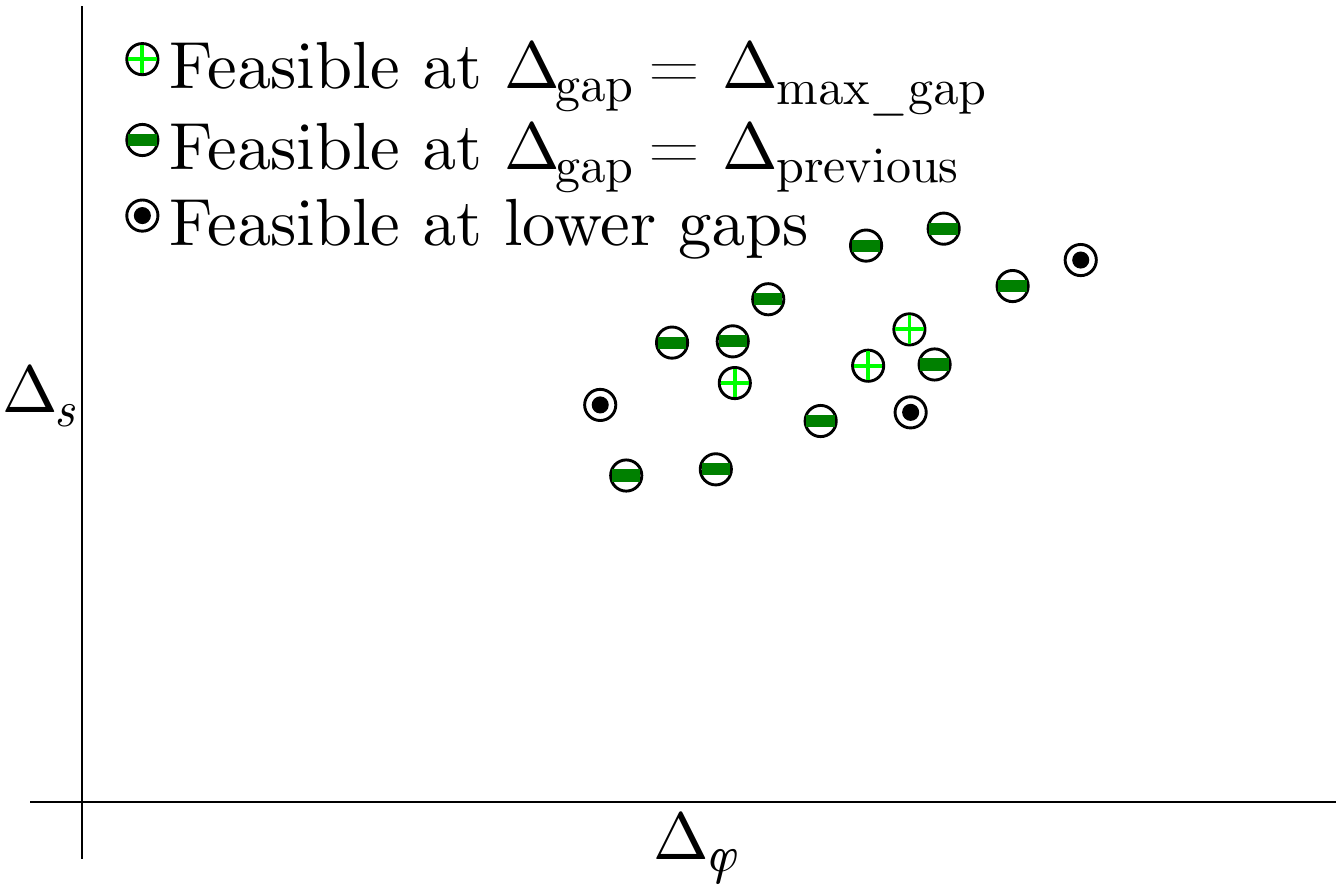}

\caption{\label{tiptop:allowed} Different types of allowed points. Only the points allowed at $\Delta_{\text{gap}}=\Delta_{\text{previous}}$ are used for rescaling.}
\end{minipage}
\end{figure}

The islands often have extreme aspect ratios in the 'natural'
coordinates.  This causes difficulties when exploring an island, so
\texttt{tiptop} rescales the coordinates.  The first step in rescaling
is to get an overall scale for all of the points (allowed,
disallowed, and in-progress) from all gaps.  We define $\Delta_{\max}$
as a scalar equal to the largest absolute coordinate value in all
dimensions, as shown in figure \ref{tiptop:box}.

For a given $\Delta_{\text{allowed}}$, we define
$\Delta_{\text{previous}}$ as the largest gap with allowed points but
less than $\Delta_{\text{allowed}}$.  This is usually a previous
value for $\Delta_{\text{allowed}}$.  Figure \ref{tiptop:allowed}
shows an example of allowed points at $\Delta_{\text{allowed}}$,
$\Delta_{\text{previous}}$, and lower gaps.

Using the $m$ allowed points at $\Delta_{\text{previous}}$, we scale the points using a
principle component analysis.  Specifically, we construct the matrix
\[
M=\left(\begin{array}{cc}
\Delta_{\phi0} & \Delta_{s0}\\
\Delta_{\phi1} & \Delta_{s1}\\
\Delta_{\phi2} & \Delta_{s2}\\
\Delta_{\phi3} & \Delta_{s3}\\
\vdots & \vdots
\end{array}\right)\,.
\]
We then compute the singular value decomposition (SVD) of this matrix
\be
M=U \Sigma V^{*}\,,
\ee
where $\Sigma$ is a rectangular $m \times {N}$ diagonal matrix with non-negative real
numbers $\sigma_{i}=\Sigma_{ii}$ on the diagonal ranging from the
smallest ($\sigma_{\min}$) to the largest ($\sigma_{\max}$).  $U$ is
an $m\times m$ unitary matrix, and $V$ is an $N\times N$ (here
$2\times 2$) unitary matrix.

We define the $N\times N$ matrix $\Omega$ as the first $N$ rows of
$\Sigma$.  This is a diagonal matrix with the entries $\sigma_{i}$, so
the inverse is trivial.  Putting this all together, we define the
rescaling matrix
\be
R \equiv \sigma_{\min} \Omega^{-1} V^{T} / \left(1.75 \times \Delta_{\max}\right)\,.
\ee
It may be that there are so few points at $\Delta_{\text{previous}}$
that they are not linearly independent.  For example, in the
beginning, there may not be any points at $\Delta_{\text{previous}}$.
If the ratio between the smallest ($\sigma_{\min}$) and largest
($\sigma_{\max}$) of these singular values is less than a tolerance
(we use $10^{-8}$), then we only scale by $\Delta_{\max}$
\be
R \equiv I / \left(1.75 \times \Delta_{\max}\right)\,,
\ee
where $I$ is the identity.

Everything is scaled by the largest coordinate value $\Delta_{\max}$
to guarantee that all points are mapped into a box with extents (-1,1)
in every dimension.  The factor of 1.75 (about $\sqrt{3}$) is to
ensure that all points will fit into the unit box even after rotation.

\begin{figure}

\begin{minipage}[t]{0.39\textwidth}%
\includegraphics[width=1\textwidth]{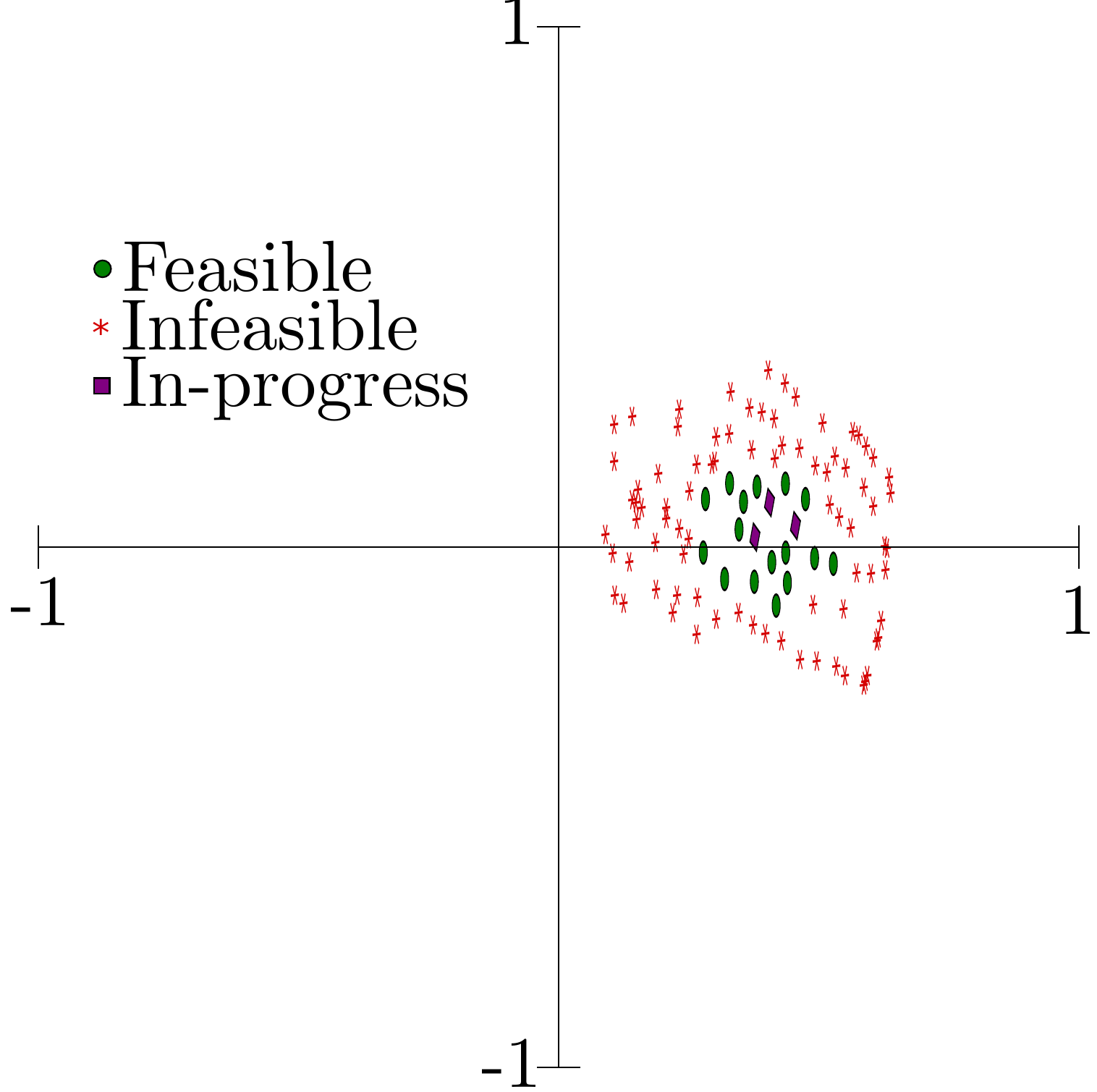}
\caption{\label{tiptop:rescaled} Points from figure \ref{tiptop:box} after rescaling.}
\end{minipage}\hfill{}%
\begin{minipage}[t]{0.59\textwidth}%
\includegraphics[width=1\textwidth]{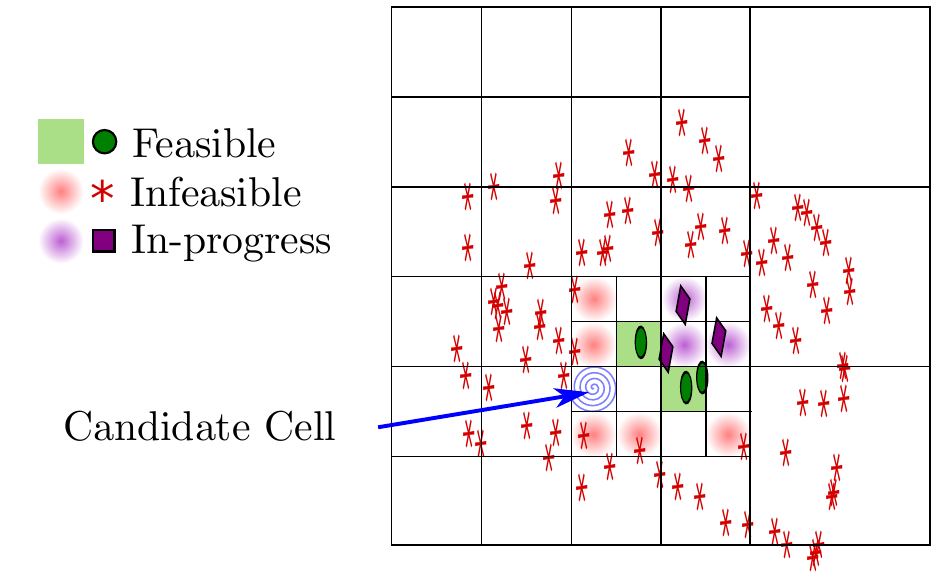}

\caption{\label{tiptop:amr} Points from figure \ref{tiptop:rescaled} with an adapted mesh.
  Points that are allowed at
    $\Delta_{\text{gap}}<\Delta_{\text{allowed}}$ have been removed.
    The blue spiral indicates an empty candidate cell.  The other
    empty cells are not diagonal from an allowed cell, so they are not
    considered.}
\end{minipage}
\end{figure}

The transformation has the effect of a rotation and then rescaling of
the rotated coordinates, so the allowed region remains convex.
However, the allowed points should outline a more circular shape than
the extended ellipse we started with, as shown in figure
\ref{tiptop:rescaled}.

One concern with this rescaling algorithm is that it weighs dense
regions with more points more than equivalently sized regions with
fewer points.  So it may not produce an optimal transform.  In
practice, the later steps spread out the points very evenly, so this
concern turns out not to be a problem in practice.

\subsubsection{Adaptively meshing the box}
\label{subsubsec:AMR}

While the distribution of points in figure \ref{tiptop:rescaled} no
longer has extreme aspect ratios, the points are still clustered in a small
region of the unit box.

Based on the assumption that the allowed island only shrinks as the
gap increases, we now only consider three sets of points: allowed at
the current $\Delta_{\text{allowed}}$, disallowed at
$\Delta_{\text{gap}} \le \Delta_{\text{allowed}}$, and in-progress.
For the rest of this step, we will be treating in-progress and
disallowed points identically.

The strategy is to place points in regions that are empty. To quantify
this emptiness, we create a hierarchy of meshes covering all the points.
Empty regions are then cells that have no points.

The mesh hierarchy starts with a single coarse cell covering the
entire unit box.  The next level has $2^N$ cells (4 for our example),
the level after that has $2^{2N}$ cells, and so on.  This subdivision
continues up to our predefined limit of 47 levels.  This implies a
relative minimum cell width of $2^{-47}$.  This is quite small, but
still significantly larger than the minimum resolution of an IEEE-754
double-precision number ($2^{-53}$).  This helps reduce errors as
points get rotated and scaled.

Next, we compute the coordinate extents of the allowed points in the
transformed frame.  For example, the extents in the transformed
horizontal direction is the difference between the smallest and
largest horizontal coordinates for the allowed points.  This gives
us a list of $N$ numbers.  In figure \ref{tiptop:rescaled}, these
extents are about 0.2 in both vertical and horizontal directions.

We then define a minimum cell size as the minimum of these $N$ extents
multiplied by a fixed fraction $f_{\text{cutoff}}$.  We use
$f_{\text{cutoff}}=2$, which is deliberately very coarse.  This tends
to double the size of the allowed region at each step.  Also, if
$f_{\text{cutoff}}$ is too small, then the algorithm will completely
fill in internal regions, even though, by assumption, the internal
spaces do not need to be checked.  This minimum cell size corresponds
to a level $l_{\max}$ in the hierarchy of meshes.

Just after a jump to a higher gap, there is only
one allowed point at $\Delta_{\text{allowed}}$.  In this case, the
extents are zero, so we set $l_{\max}=47$, the finest level.

Restricting our search to cells at level $l<l_{\max}$, we look for any
empty cells adjacent to the allowed points.  There will, in general,
be multiple empty cells at multiple levels.  We choose the largest
empty cell.

If there are multiple candidates, we choose the cell adjacent to the
first allowed point given to \texttt{tiptop}.  So when driving
\texttt{tiptop}, we always list the allowed points in the same order.
If there are multiple candidates for a single point, then we select a
new cell in the order
\begin{equation*}
  (+,+), (-,-), (+,-), (-,+).
\end{equation*}
We only check diagonals, so points get laid out in a checkerboard
pattern as in figure \ref{tiptop:amr}.

The new point is not placed at the center of the new cell, but rather
simply offset from the existing allowed point.  So if the allowed
point is in a corner of a cell, the new point will be in the same
corner of the empty cell.

In practice, the implementation does not explicitly create the mesh at
all levels.  Rather, the points are stored in a tree.  A node in the
tree can have up to $2^{N}$ leaves, but leaves are only created if there
is a point in that leaf.  Adding a point to the tree adds it at all
levels, so that it is easy to determine if a cell is occupied at any
level.

The observed behavior of this algorithm is that it quickly finds a
rough estimate for the boundary between allowed and disallowed, but
can spend a lot of effort finding the exact boundaries.  In-progress
points are treated as disallowed, so too many in-progress points will
lead to extra work.  In practice, we have up to 16 points in-progress
at any one time.

\subsection{Jumping to a larger gap}
\label{subsec:jumping}

If the previous section does not yield a new point, and there are no
in-progress points at that gap, then we try to jump to a larger gap.

We start by rescaling the points as in section
\ref{subsubsec:Rescaling}.  We draw a coordinate box around all of the
points allowed at $\Delta_{\text{allowed}}$ and shrink it by a
factor of 2.  Then we find the largest gap $\Delta_{\text{ceiling}}$
that can accommodate this box without containing any disallowed or
in-progress points with
$\Delta_{\text{gap}} \le \Delta_{\text{ceiling}}$.  At the beginning,
there are no disallowed points at large gaps, so
$\Delta_{\text{ceiling}} = \Delta_{\text{max\_gap}}.$

Defining $\Delta_{\text{diff}} \equiv \Delta_{\text{ceiling}} - \Delta_{\text{allowed}}$, 
we return the center of the box at
$\Delta_{\text{gap}}=\Delta_{\text{allowed}} +
  \Delta_{\text{diff}}/2,$ thus bisecting the range of
allowed gaps.  This underscores the need for a good estimate of
$\Delta_{\text{max\_gap}}.$ If the estimate is too high, then the
algorithm will recommend too many points that are far too large.

One thing to note is that when running with multiple in-progress
points, each subsequent point will be at a smaller
$\Delta_{\text{gap}}$.  So if there are 3 points running concurrently,
they will be placed at
$\Delta_{\text{allowed}} + \Delta_{\text{diff}}/2$,
$\Delta_{\text{allowed}} + \Delta_{\text{diff}}/4$, and
$\Delta_{\text{allowed}} + \Delta_{\text{diff}}/8$.  If the first
point at $\Delta_{\text{allowed}} + \Delta_{\text{diff}}/2$ succeeds,
then any effort towards verifyinig the latter two points will be
wasted.

This partition procedure means that a rough estimate for the upper
bound is $\Delta_{\text{ceiling}}.$ This assumes both that the island
itself is convex and that the allowed region near the tip of the
allowed region is convex, as in Figure \ref{tipfig:islands2d}.

Overall, we have found this approach to work reasonably well.  More
importantly, it is very robust.  It is very easy to be too clever,
resulting in odd failures.

\section{Results}
\label{sec:results}

\subsection{Dimension bounds with OPE scans}
\label{sec:dimisland}
Next we present our conformal bootstrap island computed using \texttt{SDPB} \cite{Simmons-Duffin:2015qma,Landry:2019qug}, along with its comparison with various Monte Carlo results.  Computing the conformal bootstrap island requires scanning over the three operator dimensions $\{\De_\f,\De_s,\De_t\}$ using the Delaunay search algorithm described in~\cite{Chester:2019ifh}, and for each point using the ``cutting surface" algorithm presented in~\cite{Chester:2019ifh} to decide if there exists an allowed point in the space of OPE coefficient ratios $\{\frac{\l_{sss}}{\l_{\f\f s}}, \frac{\l_{tts}}{\l_{\f\f s}}, \frac{\l_{\f\f t}}{\l_{\f\f s}}, \frac{\l_{ttt}}{\l_{\f\f s}}\}$.

When computing the island we make the following assumptions about the spectrum unless stated otherwise. We assume that $\phi$, $s$, and $t$ are the only relevant operators in their respective symmetry representations, so that $\Delta_{\phi',s',t'} \geq 3$. In addition, we assume that the leading rank-4 scalar has a dimension satisfying $\Delta_{t_4} \geq 2$. We assume an \Othree current with $\Delta_{J} = 2$ and stress tensor with $\Delta_{T} = 3$, with coefficients satisfying the Ward identity constraints. We also impose a twist gap above them, as well as in all other sectors not mentioned above, of size $10^{-6}$.

\begin{figure}[htbp]
  \centering
  \includegraphics[width=0.85\textwidth]{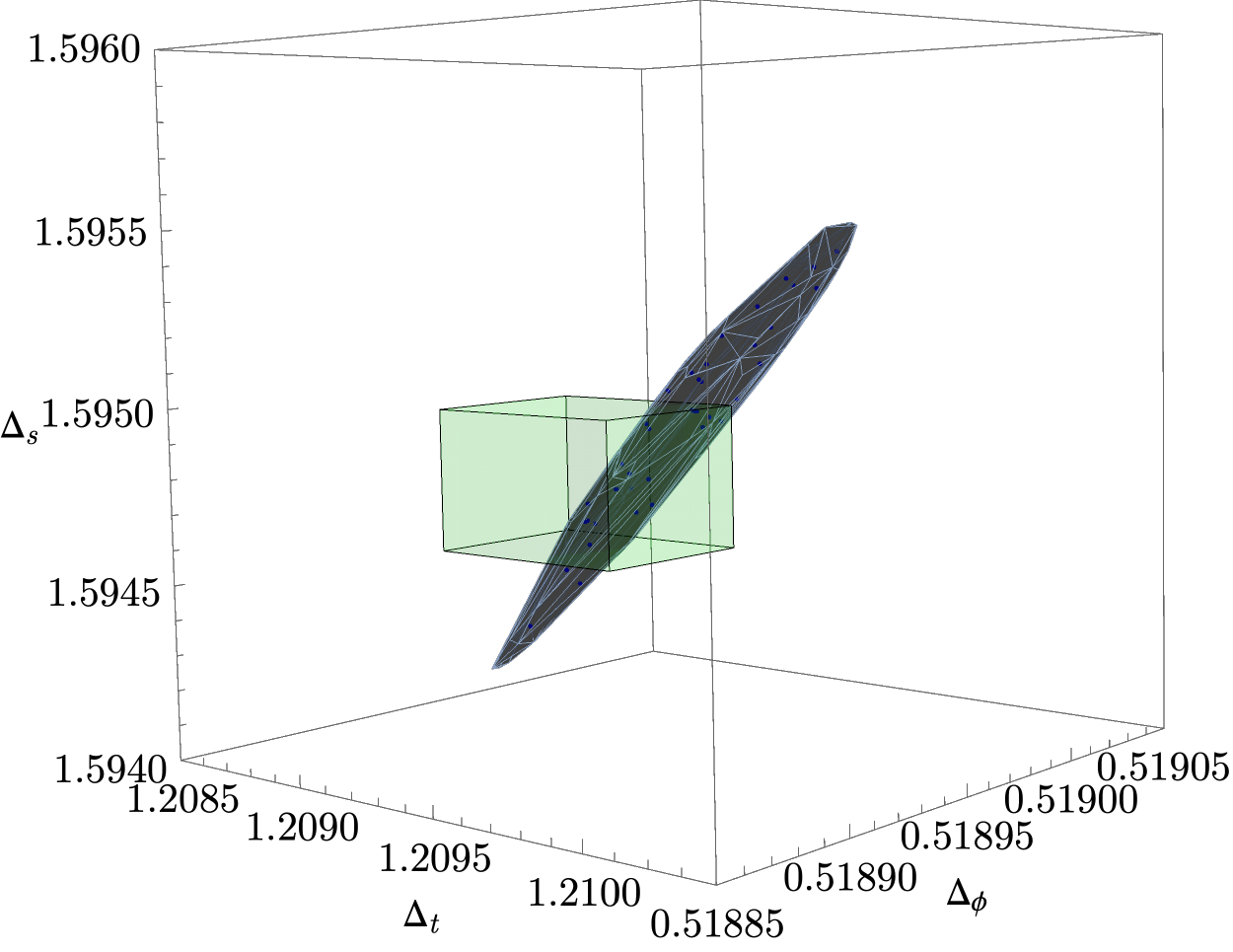}
  \caption{\label{dimension:only22} The $\Lambda=43$ conformal bootstrap dimension island (black) compared with the Monte Carlo results \cite{Hasenbusch:2020pwj,Hasenbusch_2011} (green).}
\end{figure}

\begin{figure}[htbp]
  \centering
  \includegraphics[width=0.7\textwidth]{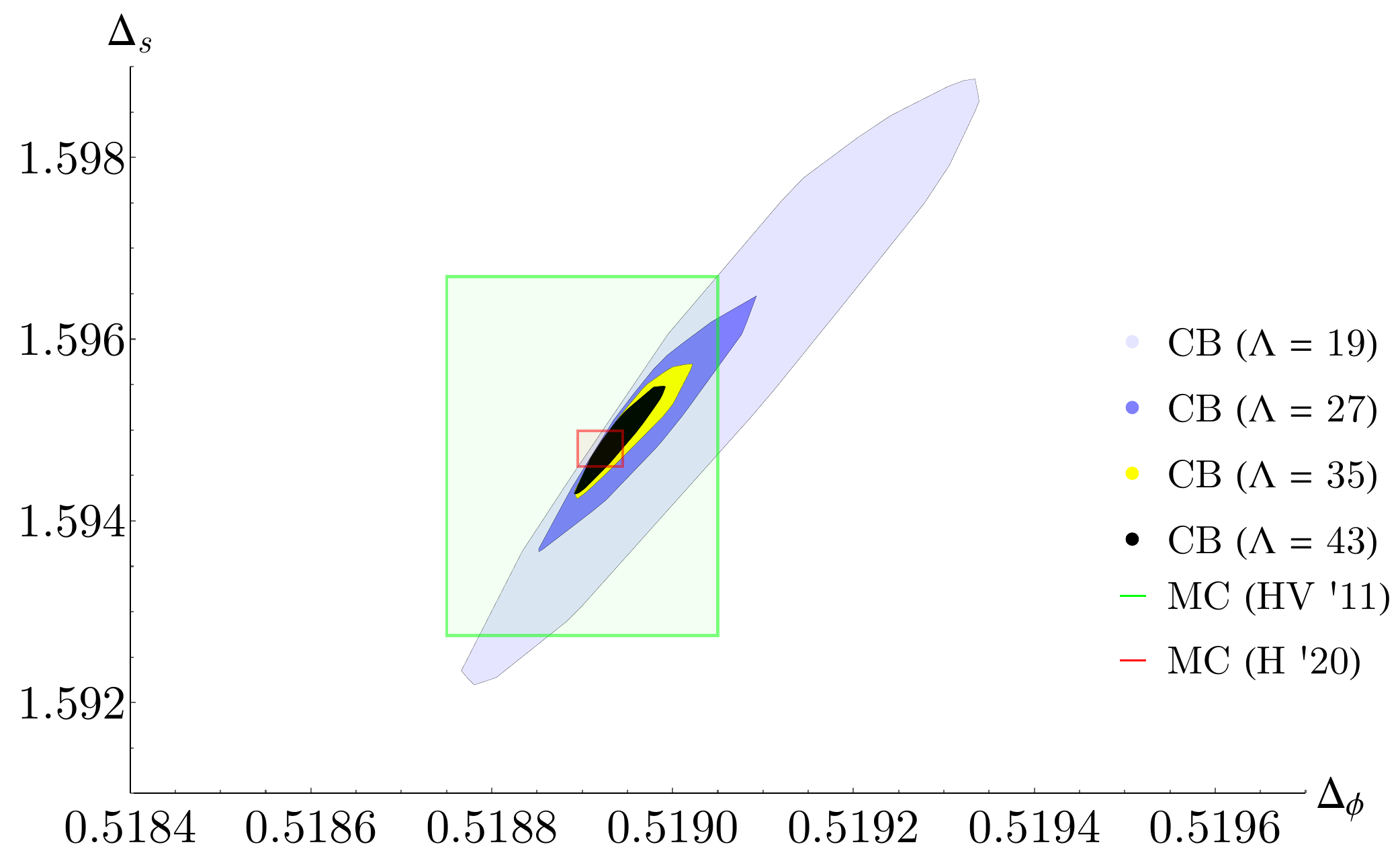}
    \includegraphics[width=0.7\textwidth]{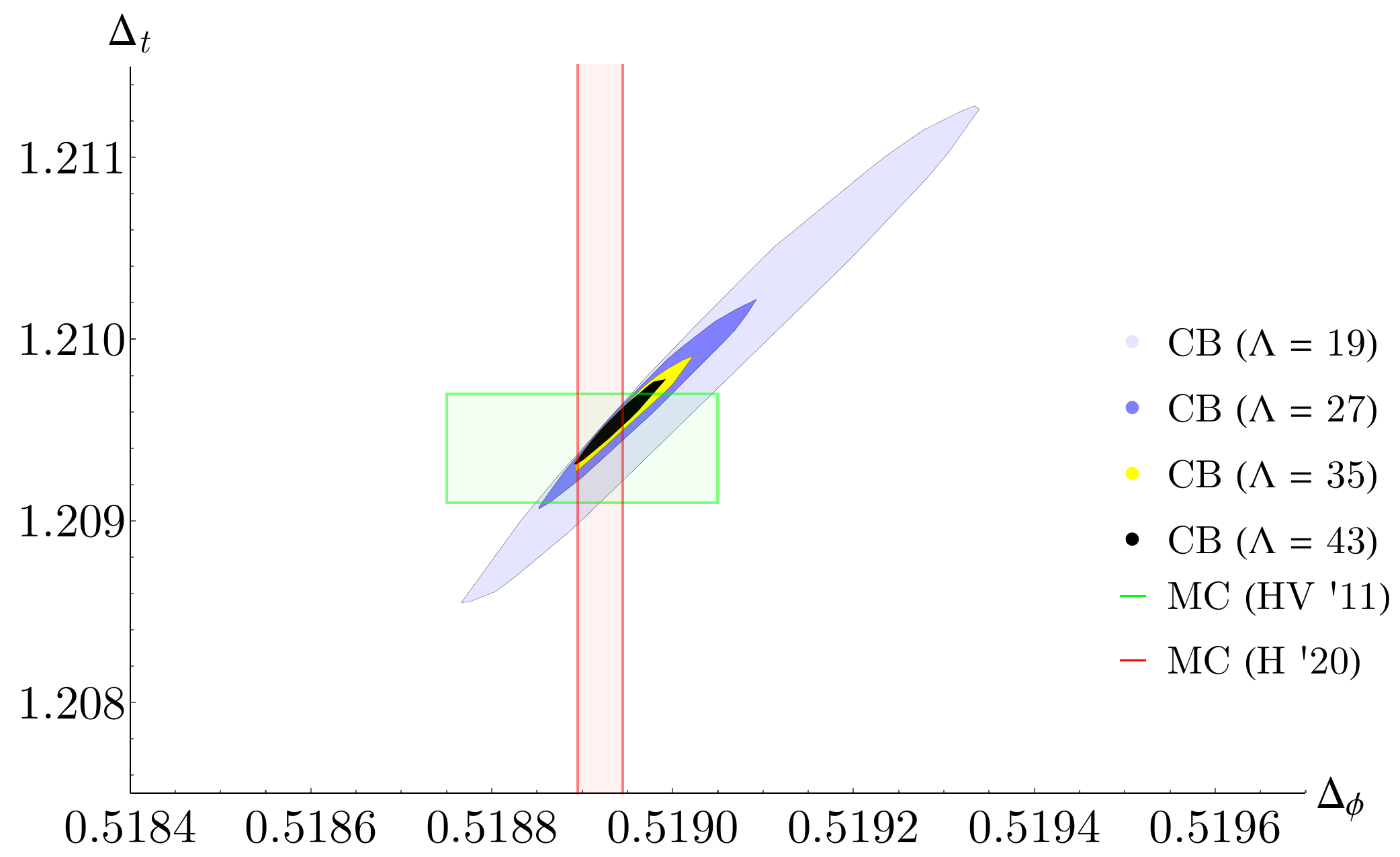}
      \includegraphics[width=0.7\textwidth]{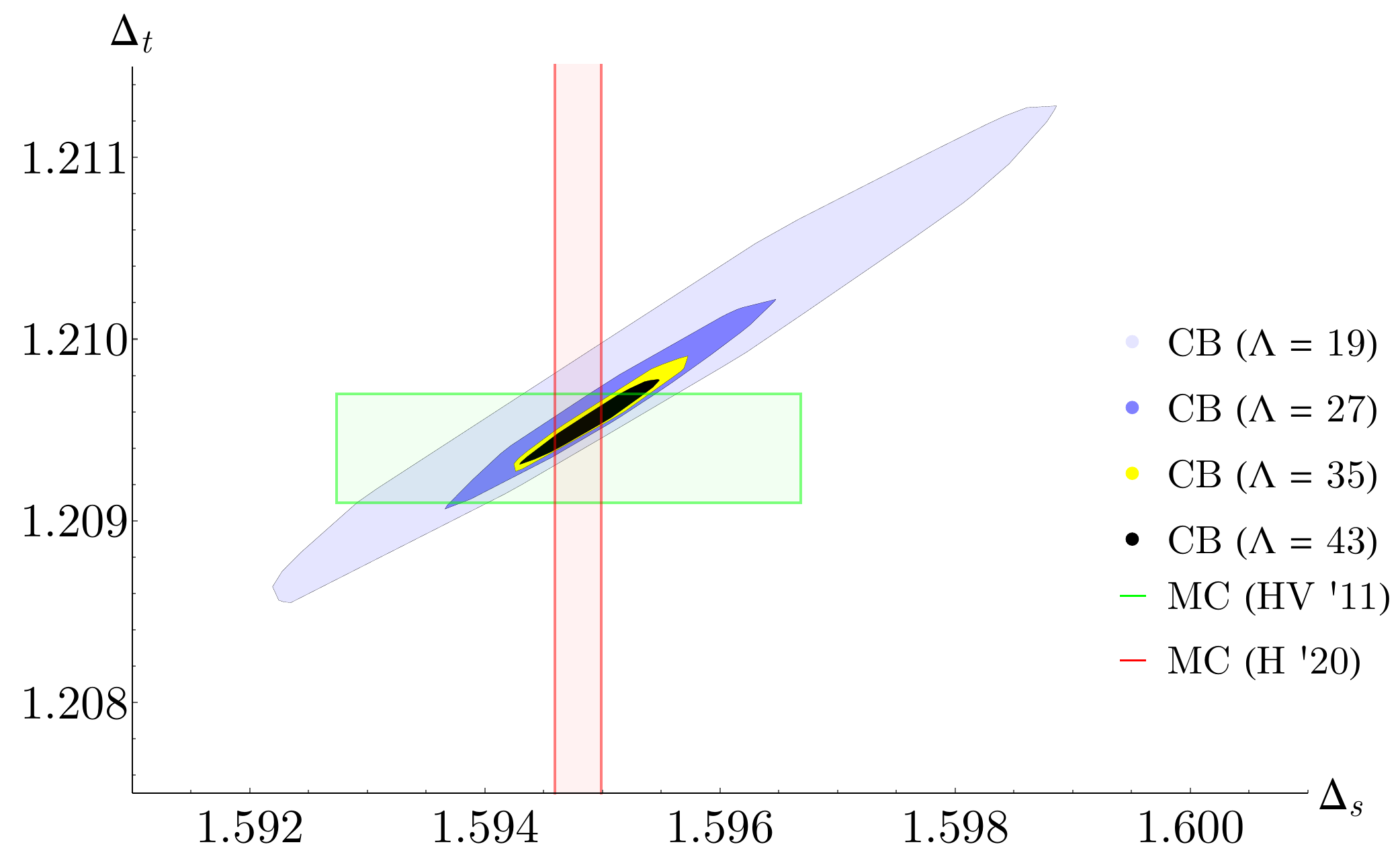}
  \caption{\label{dimension:2d} Comparison between the conformal bootstrap islands at $\Lambda=19,27,35,43$ projected to the $\{\Delta_{\phi}, \Delta_{s}\}$, $\{\Delta_{\phi}, \Delta_{t}\}$, and $\{\Delta_{s}, \Delta_{t}\}$ planes and the Monte Carlo results of~\cite{Hasenbusch:2020pwj,Hasenbusch_2011}.}
\end{figure}

In figure~\ref{dimension:only22} we show the conformal bootstrap island we have computed at $\Lambda=43$ using these assumptions, compared to the Monte Carlo results of~\cite{Hasenbusch:2020pwj,Hasenbusch_2011}. In figure~\ref{dimension:2d} we show various 2d projections of the bootstrap island. In appendix~\ref{app:points} we give the full set of allowed and disallowed points we computed at $\Lambda=43$, along with figure~\ref{dimension:affine} which shows the convergence of the allowed points as a function of $\Lambda$ after performing an affine transformation to make the allowed regions roughly spherical.

In these plots we show our best determination of the allowed region at a given $\Lambda$, constructed by computing a Delaunay triangulation of the tested points, choosing triangles that contain both allowed and disallowed points, and plotting the convex hull of the points that are midway between the allowed and disallowed vertices in these triangles. At $\Lambda=43$, this ``best-fit" region gives
\begin{align}
\Delta_{\phi} &= 0.518942(51^*)\,,\nn\\
\Delta_{s} &= 1.59489(59^*)\,,\nn\\
\Delta_{t} &= 1.20954(23^*)\,.
\end{align}
A more rigorous determination can be made by taking the convex hull of the disallowed points in these boundary Delaunay triangles. This region gives the rigorous error bars
\begin{align}
\Delta_{\phi} &= 0.518936({\bf 67})\,,\\
\Delta_{s} &= 1.59488({\bf 81})\,,\\
\Delta_{t} &= 1.20954({\bf 32})\,,
\end{align}
which we have quoted in table~\ref{tab:results}. 

The allowed points at $\Lambda=43$ are associated with OPE coefficient ratios which live in the ranges\footnote{Note that there is an ambiguity in the signs of these coefficients, related to performing the operator redefinitions $s \rightarrow -s$ and $t \rightarrow -t$. This freedom can be used to fix $\lambda_{\f\f s}$ and $\lambda_{\f\f t}$ to be positive, after which all other signs in (\ref{eq:OPEs}) are determined to be positive by the conformal bootstrap. }
\begin{align}
\frac{\lambda_{sss}}{\lambda_{\f\f s}} &= 0.9643(20^*)\,,\nn\\
\frac{\lambda_{tts}}{\lambda_{\f\f s}} &= 1.87593(53^*)\,,\nn\\
\frac{\lambda_{\f\f t}}{\lambda_{\f\f s}} &= 1.66808(23^*)\,,\nn\\
\frac{\lambda_{ttt}}{\lambda_{\f\f s}} &= 2.86034(61^*)\,.
\label{eq:OPEs}\end{align}
These should be viewed as an approximation to the full allowed region of OPE coefficients, which may be slightly larger.

\subsection{Central charges and $\lambda_{\phi \phi s}$}\label{optimal}

Next, we compute upper and lower bounds on the magnitude of the OPE coefficient $\lambda_{\f\f s}$, the central charge $C_T$, and the current central charge $C_J$. We compute these bounds over a small sample of points in our allowed region so the results will be inherently non-rigorous. However, we believe that this treatment gives reasonable estimates for these quantities that are more precise than previous results. 

The strategy is similar to the method we employed in~\cite{Chester:2019ifh}. We take seven primal points in the $\Lambda=43$ island, consisting of the scaling dimensions and allowed OPE coefficients. The points are chosen to be sufficiently symmetrized and sparse across the $\Lambda=43$ island we have computed. For each of these points, we extremize $C_T$, $C_J$, and the external OPE norm parameterized by $\lambda_{\phi \phi s}$, to obtain upper and lower bounds. This calculation was limited to $\Lambda=35$ due to our available computational resources. The data points and \texttt{SDPB} parameters we used are summarized in tables \ref{tab:the7points} and \ref{extremalparameter}, respectively. 

There is an important comment we want to make about the upper bound computation on $C_T$ and $C_J$ (a similar comment was made in \cite{Chester:2019ifh}). For computing upper bounds on $C_T$ and $C_J$, we have to assume a gap $\Delta_{T'/J'}$ above the unitarity bound for the next operators in the $T$ or $J$ sectors. Note that this gap was not assumed in our OPE scan, so this extra constraint might turn an allowed point into a disallowed point. If we do not have such a gap, the upper bound is loose and may not give reasonable results. On the other hand, large gaps can make \texttt{SDPB} unable to find a solution. 

In table \ref{tab:the7points}, we summarize the gaps $\Delta_{T'/J'}$ we assume in the upper bound calculations. From spectrum determinations using the extremal functional method (see \cite{Poland:2010wg,ElShowk:2012hu}), we have noticed that a gap $\Delta_{T'/J'}=\Delta_{T/J} + 1$ above $T$ and $J$ is generally favored. We were able to compute bounds with this gap for three of the points, but for the other four we could not find solutions.  For those points, we adopted the weaker assumption $\Delta_{\text{ext},T(J)}=\Delta_{T/J}+0.1$.

Following this procedure, we obtain our estimates of $C_T$, $C_J$, and $\lambda_{\phi \phi s}$ in the critical \Othree model,
\begin{align}
&{C_J}/C_J^{{\rm{free }}} = {\rm{0.90632(}}{{\rm{16}}^*}{\rm{)}}\,,\nonumber\\
&{C_T}/C_T^{{\rm{free }}} = {\rm{0.944524(}}{{\rm{28}}^*}{\rm{)}}\,,\nonumber\\
&{\lambda _{\phi \phi s}} = {\rm{0.524261(}}{{\rm{59}}^*}{\rm{)}}\,.
\end{align}
These results agree with and are more precise than previous determinations of these quantitites (see \cite{Kos:2013tga,Kos:2015mba,Kos:2016ysd}). 

\subsection{Upper bound on $\Delta_{t_4}$}
\begin{figure}[ht]
  \centering
      \includegraphics[width=0.75\textwidth]{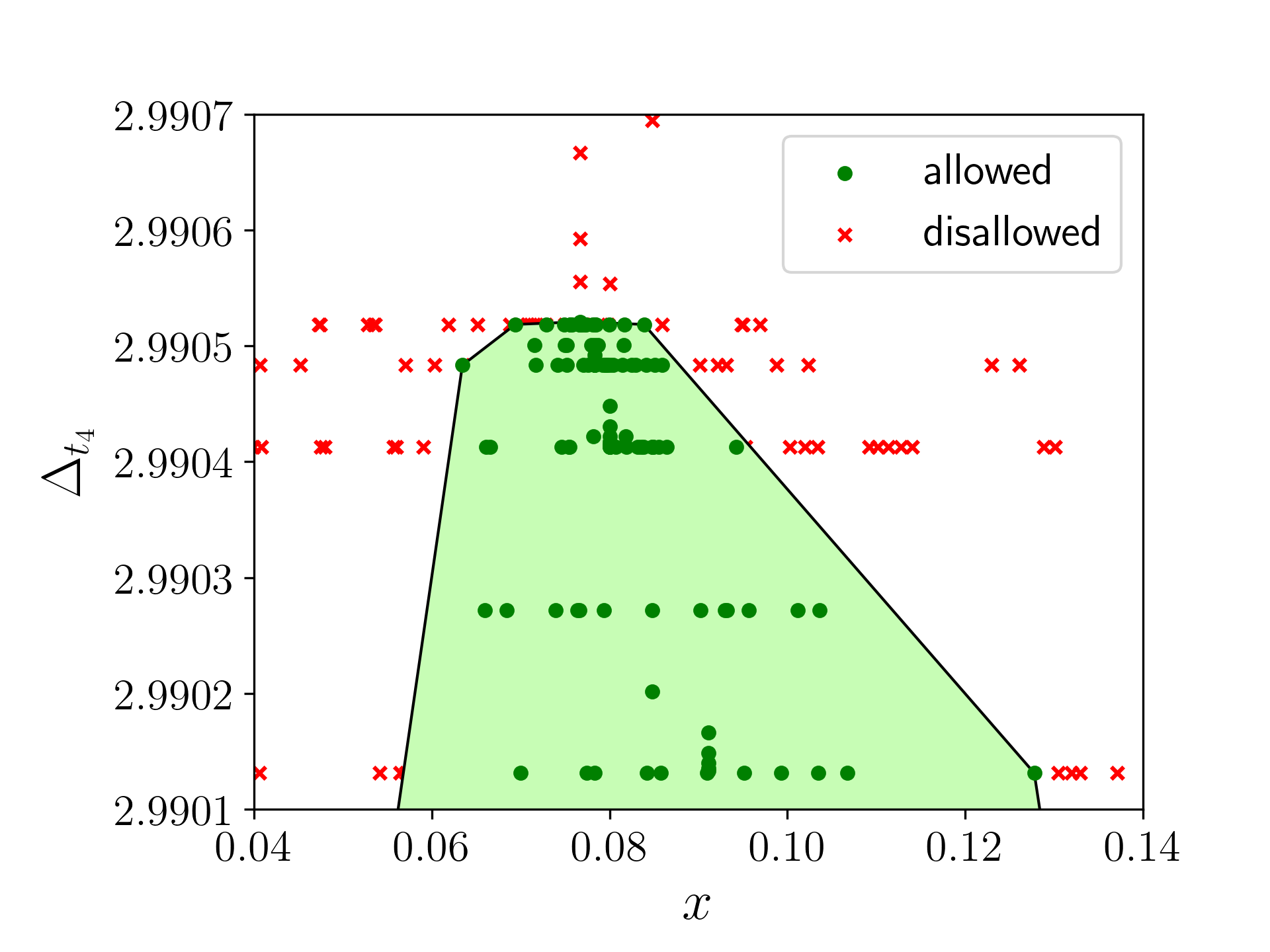}
  \caption{\label{tipfig:islands2d} Two-dimensional projection of the results of the \texttt{tiptop} search at $\Lambda=35$. The $x$ coordinate is related to the three scalar dimensions via~(\ref{eq:affine}). Projections in $y$ and $z$ look similar.  We have superimposed a convex hull encompassing the allowed points on top, obscuring some of the disallowed points.  We can see the behaviour of the \texttt{tiptop} algorithm, exploring the island at one $\Delta_{t_4}$ before jumping to a larger $\Delta_{t_4}$.  The jumps become progressively smaller, indicating convergence.  We computed 16 points simultaneously, and this calculation took several months during which the \texttt{tiptop} algorithm was being developed.  So the points reflect occasional crashes and small inefficiencies in the set of computed points.}
\end{figure}

\begin{figure}[ht]
  \centering
  \includegraphics[width=0.75\textwidth]{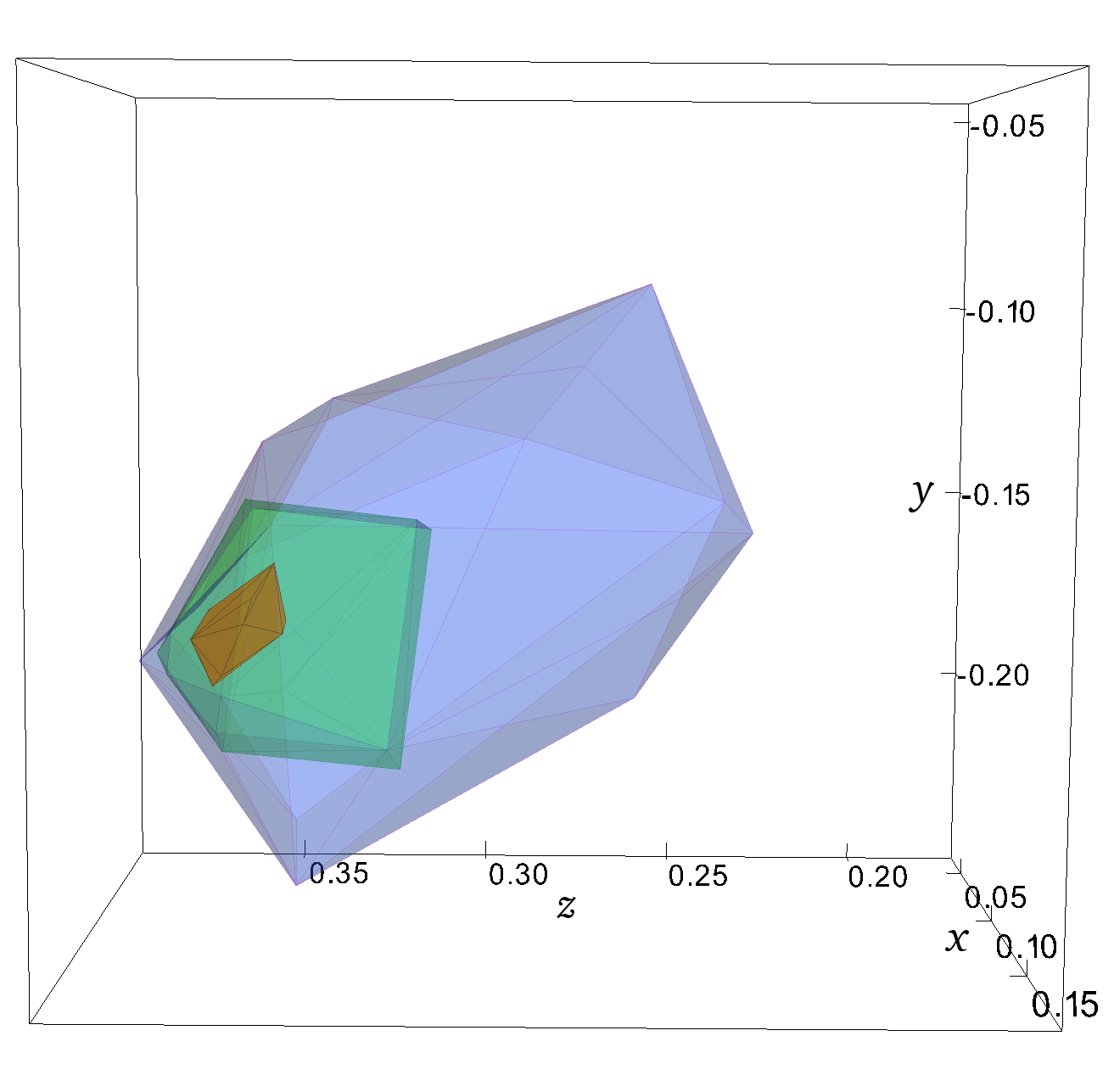}
  \caption{\label{tipfig:islands3d} Three-dimensional islands of allowed points at different $\Delta_{t_4}$ at $\Lambda=35$, demonstrating how the islands shrink as we approach the maximum $\Delta_{t_4}$. The $x$, $y$, and $z$ coordinates are related to the three scalar dimensions via~(\ref{eq:affine}). The values for $\Delta_{t_4},$ from the largest region to smallest, are $2.989$, $2.99025$, and $2.9905$, with smaller values including all allowed points at larger values.}
\end{figure}

Our last result is the maximum value of the rank-4 scalar dimension
$\Delta_{t_4}$. In conjunction with the \texttt{tiptop} algorithm
described in section~\ref{sec:tiptop}, we computed points at
$\Lambda=19,$ $27,$ and $35$.  Allowed points at lower values of
$\Lambda$ were used to initiate the search at larger values of
$\Lambda$.  Figure \ref{tipfig:islands2d} shows a projection of a
subset of the 1311 disallowed points and 172 allowed points at
$\Lambda=35$, and Figure \ref{tipfig:islands3d} shows how the island
shrinks as we approach the maximum $\Delta_{t_4}$.
 
The largest allowed value of $\Delta_{t_4}$ was a single point at
$2.99052$ with the scaling dimensions
$\{\Delta_{\phi}, \Delta_{s}, \Delta_t\} = \{0.518962, 1.59527,
1.20969\}$.  The limit from $\Delta_{\text{ceiling}}$ (see Section
\ref{subsec:jumping}) is
\begin{align}
\Delta_{t_4} < 2.99056\,. 
\end{align}
This implies that the leading rank-4 tensor in the critical \Othree model is relevant, in agreement with other studies.

\section{Future directions}
\label{sec:conclusions}

In this work we have applied the methods developed in~\cite{Chester:2019ifh,Liu:2020tpf} for large-scale bootstrap problems to the critical \Othree model in three dimensions. This has led to results for scaling dimensions which are competitive with the most precise Monte Carlo simulations, and results for OPE coefficients which are significantly more precise than previous determinations. In addition, we have computed a rigorous bound on the scaling dimension of the leading rank-4 tensor, showing that it is relevant. Thus, any \Othree system with cubic anisotropy should flow to the cubic fixed point (discussed in section~\ref{sec:ONvscubic}) instead of the Heisenberg fixed point.

An interesting direction for future research will be the application of conformal perturbation theory to this flow. The cubic model can be reached by perturbing the \Othree CFT with the operator $X\equiv\sum_{i=1}^3 t_4^{iiii}$, which breaks \Othree symmetry to the discrete symmetry $\mathds Z_3 \rtimes \mathcal S_3$. From the \Othree point of view, this term is a certain component of the \Othree rank-4 tensor with dimension \(\Delta _{t_4}\simeq 2.99\).
On the other hand, in the cubic fixed point conformal perturbation theory predicts \(\Delta _X\simeq 3.01\). Because this term is marginally irrelevant with \(\delta = \Delta_{X} - 3 \simeq 0.01\), if we want to reach the cubic fixed point by a Monte Carlo simulation, the size of the lattice has to be around the order of \(2^{1/\delta }\), which is impractical to implement. 

An alternative way to estimate the cubic CFT data is using conformal perturbation theory. We start with the perturbed action $S=S_{\Othreewithout}+g\int d^3x X$. Using the formalism in \cite{Komargodski:2016auf}, one finds the beta function to be
\begin{equation}
\beta _g=-\delta  g-\frac{\vol\, S^{d-1}}{2} \lambda_{XXX} g^2.
\end{equation}
The dimension of an operator \(\mathcal{O}\) at the cubic fixed point is then given at linear order in $\de$ by \(\Delta _{\mathcal{O} }=\Delta _0+2\delta \lambda_{\mathcal{O}\mathcal{O} X}/\lambda_{XXX}\), where $\Delta _0$ is the dimension of corresponding operator in the \Othree CFT. Specifically one obtains \(\Delta _X=\Delta _0+2\delta\), which justifies the estimate \(\Delta _X\simeq 3.01\).

The OPE coefficient \(\lambda_{XXX}\) is proportional to \(\lambda_{t_4t_4t_4}\). Unfortunately, using the setup of the present paper, we don{'}t have access to \(\lambda_{t_4t_4t_4}\). To access \(\lambda_{t_4t_4t_4}\), one needs to
bootstrap all four-point functions involving $\{\phi ,s,t,t_4\}$, which is a concrete task for future research. Here we can estimate that the correction to \(\Delta
_t\) in the cubic fixed-point is of order \(\delta =0.01\). On the other hand, the corrections to \(\Delta _{\phi },\Delta _s\) start at order \(\delta ^2\simeq 0.0001\) since $\lambda_{\phi\phi t_4}=\lambda_{sst_4}=0$.
Note that in this work the error bar for \(\Delta _t\) is much smaller than $\delta $. Therefore a careful study of the \(\{\phi ,s,t,t_4\}\) system should yield a solid prediction for the correction to \(\Delta _t\) in the cubic model. 

Of course, it will also be interesting to understand how to isolate the cubic fixed point more directly using the conformal bootstrap, perhaps using a larger system of correlators than was considered in~\cite{Rong:2017cow, Kousvos:2018rhl,Kousvos:2019hgc}. One can also straightforwardly apply the large-scale bootstrap techniques we have developed to other \ON models, as well as to 3d CFTs with fermions (using the newly developed software~\cite{Erramilli:2020rlr}) or to study conserved currents \cite{Dymarsky:2017xzb,Dymarsky:2017yzx,Reehorst:2019pzi}. Using these methods one can also continue exploring larger systems of correlators that may help us to isolate CFTs containing gauge fields, such as 3d QED \cite{Chester:2016wrc,Chester:2017vdh} and 4d QCD. 

Now that we have precisely isolated the \Othree model, we are also in position to do a more detailed study of its low-twist trajectories of operators as a function of spin, which can be compared to analytical calculations using the Lorentzian Inversion formula~\cite{Caron-Huot:2017vep,Simmons-Duffin:2017nub}, following the approach of~\cite{Albayrak:2019gnz,Liu:2020tpf,Caron-Huot:2020ouj}. Such analytical techniques can also be used to estimate the leading Regge intercepts and related Lorentzian data of the \Othree model. In future work it will also be important to understand how to better incorporate insights from the analytical bootstrap, such as our precise understanding of the large spin asymptotics, into making large-scale numerical methods even more powerful.

\section*{Acknowledgements}
We thank Yinchen He, Igor Klebanov, Filip Kos, Zhijin Li, Jo\~ao Penendones, Junchen Rong, Slava Rychkov, Andreas Stergiou, and Ettore Vicari for discussions. WL, JL, and DSD are supported by Simons Foundation grant 488657 (Simons Collaboration on the Nonperturbative Bootstrap). DSD and JL are also supported by a DOE Early Career Award under grant no. DE-SC0019085. DP is supported by Simons Foundation grant 488651 (Simons Collaboration on the Nonperturbative Bootstrap) and DOE grants DE-SC0020318 and DE-SC0017660. This project has received funding from the European Research Council (ERC) under the European Union's Horizon 2020 research and innovation programme (grant agreement no.~758903). AV is also supported by the Swiss National Science Foundation (SNSF) under grant no.~PP00P2-163670. SMC is supported by a Zuckerman STEM Leadership Fellowship.

This work used the Extreme Science and Engineering Discovery Environment (XSEDE) Comet Cluster at the San Diego Supercomputing Center (SDSC) through allocation PHY190023, which is supported by National Science Foundation grant number ACI-1548562. This work also used the EPFL SCITAS cluster, which is supported by the SNSF grant PP00P2-163670, the Caltech High Performance Cluster, partially supported by a grant from the Gordon and Betty Moore Foundation, and the Grace computing cluster, supported by the facilities and staff of the Yale University Faculty of Sciences High Performance Computing Center.

\appendix
\section{Code availability}
\label{tiptop:availability}

All code used in this work is available online. This includes the various codes described in appendix A of~\cite{Chester:2019ifh}, as well as
\texttt{tiptop}, available at 
  \url{https://gitlab.com/bootstrapcollaboration/tiptop}.
\texttt{tiptop} is implemented in C++17 and uses the Boost
\cite{boost}, Eigen \cite{eigenweb}, and VTK \cite{vtk} libraries.  The version used in this paper has the Git commit hash
\begin{lyxcode}
23774017b8726699bd838cf138a65e29405f0907
\end{lyxcode}

\section{Software setup and parameters}

The computations of the \Othree model islands described in section~\ref{sec:dimisland} with $\Lambda = 19,27$ were performed on the Caltech HPC Cluster, the Yale Grace Cluster, and the EPFL SCITAS cluster. For the computations with $\Lambda = 35,43$, we tested possible primal points using the Caltech and Yale clusters. After finding a few initial primal points, the main Delaunay triangulation search was performed on the XSEDE Comet Cluster~\cite{6866038} at the San Diego Supercomputing Center through allocation PHY190023. Together, the computations of the $\Lambda=35$ island, the $\Lambda=43$ island, and the $\Lambda=35$ {\tt tiptop} search took 2.94M CPU hours on the Comet Cluster. The optimization computations of section~\ref{optimal} were performed at $\Lambda=35$ and completed on the Caltech and Yale clusters.

We used the following choices for the set of spins at each value of $\Lambda$:
\begin{align}\label{eq:spinsets}
S_{19} &= \{0,\dots,26\}\cup \{49,50\}\,,\nonumber\\
S_{27} &= \{0,\dots,31\}\cup \{49,50\}\,,\nonumber\\
S_{35} &= \{0,\dots,44\}\cup \{47, 48, 51, 52, 55, 56, 59, 60, 63, 64, 67, 68\}\,,\nonumber\\
S_{43} &= \{0,\dots,64\}\cup \{67, 68, 71, 72, 75, 76, 79, 80, 83, 84, 87, 88\}\,.
\end{align}
The \texttt{SDPB} parameters used in our computations are given in tables~\ref{tab:params} and~\ref{tab:optparams}. 

\begin{table}
\begin{center}
\begin{tabular}{@{}c|c|c|c|c@{}}
	\toprule
$\Lambda$ &  19 & 27 & 35 & 43 \\
{\small\texttt{keptPoleOrder}}& 14 & 14 & 32 & 40 \\
{\small\texttt{order}}& 60  & 60 & 80 & 90 \\
{\small\texttt{spins}} & $S_{19}$ & $S_{27}$ & $S_{35}$ & $S_{43}$  \\
{\small\texttt{precision}} & 768 & 768 & 960& 1024 \\
{\small\texttt{dualityGapThreshold}} & $10^{-30}$ & $10^{-30}$ & $10^{-30}$  & $10^{-75}$ \\
{\small\texttt{primalErrorThreshold}}& $10^{-200}$ & $10^{-200}$ & $10^{-200}$ & $10^{-200}$ \\
{\small\texttt{dualErrorThreshold}} & $10^{-200}$ & $10^{-200}$ & $10^{-200}$ & $10^{-200}$\\ 
{\small\texttt{initialMatrixScalePrimal}} & $10^{40}$ & $10^{50}$& $10^{50}$ & $10^{60}$\\
{\small\texttt{initialMatrixScaleDual}} & $10^{40}$ & $10^{50}$& $10^{50}$ & $10^{60}$\\
{\small\texttt{feasibleCenteringParameter}} & 0.1 & 0.1 & 0.1 & 0.1\\
{\small\texttt{infeasibleCenteringParameter}} & 0.3 & 0.3 & 0.3 & 0.3\\
{\small\texttt{stepLengthReduction}} & 0.7 & 0.7 & 0.7 & 0.7\\
{\small\texttt{maxComplementarity}} & $10^{100}$ & $10^{130}$ & $10^{160}$ & $10^{200}$\\
 \bottomrule
\end{tabular}
\caption{\label{tab:params}Parameters used for the computations of the conformal bootstrap islands in section~\ref{sec:dimisland}. The sets $S_{\Lambda}$ are defined in (\ref{eq:spinsets}).}
\end{center}
\end{table}

\begin{table}
\begin{center}
\begin{tabular}{@{}c|c@{}}
	\toprule
$\Lambda$ &  35 \\
{\small\texttt{keptPoleOrder}}&  30 \\
{\small\texttt{order}}&  60 \\
{\small\texttt{spins}} & $S_{35}$ \\
{\small\texttt{precision}} &  768 \\
{\small\texttt{dualityGapThreshold}} & $10^{-20}$ \\
{\small\texttt{primalErrorThreshold}}& $10^{-50}$ \\
{\small\texttt{dualErrorThreshold}} & $10^{-60}$\\ 
{\small\texttt{initialMatrixScalePrimal}} & $10^{50}$\\
{\small\texttt{initialMatrixScaleDual}} & $10^{50}$\\
{\small\texttt{feasibleCenteringParameter}} & 0.1\\
{\small\texttt{infeasibleCenteringParameter}} & 0.3\\
{\small\texttt{stepLengthReduction}} & 0.7\\
{\small\texttt{maxComplementarity}} & $10^{160}$\\
 \bottomrule
\end{tabular}
\caption{\label{tab:optparams}Parameters used for the optimization computations in section~\ref{optimal}. The set $S_{35}$ is defined in (\ref{eq:spinsets}).}
\label{extremalparameter}
\end{center}
\end{table}

\section{Tensor structures}
\label{tensApp}

In this appendix we compute the \SOthree tensor structures $T^\mathcal{R}$ that appear in the conformal block expansions \eqref{4point} for the 4-point functions listed in table \ref{table}. We start by defining a basis of tensors for each configuration in table \ref{table}:
\es{bases}{
   B_{\phi\phi\phi\phi}^I&=\begin{pmatrix} \delta_{ij}\delta_{kl} \\  \delta_{ik}\delta_{jl} \\  \delta_{il}\delta_{jk} \end{pmatrix}\,,\\
B^I_{tttt}&=\begin{pmatrix} \delta_{i_1 j_1}\delta_{i_2 j_2}\delta_{k_1 l_1}\delta_{k_2 l_2} \\  
\delta_{i_1 k_1}\delta_{i_2 k_2}\delta_{j_1 l_1}\delta_{j_2 l_2}\\ 
 \delta_{i_1 l_1}\delta_{i_2 l_2}\delta_{k_1 j_1}\delta_{k_2 j_2}\\ 
  \delta_{i_1 j_1}\delta_{i_2 k_2}\delta_{k_1 l_1}\delta_{j_2 l_2}\\ 
   \delta_{i_1 j_1}\delta_{i_2 l_1}\delta_{j_2 k_1}\delta_{k_2 l_2} \end{pmatrix}\,,\\
B_{t\phi t\phi}^I&=\begin{pmatrix} \delta_{i_1 k_1}\delta_{i_2 k_2}\delta_{j_1 l_1}\\  
\delta_{i_1 j_1}\delta_{i_2 k_2}\delta_{k_1 l_1}\\ 
 \delta_{i_1 l_1}\delta_{i_2 k_2}\delta_{j_1 k_1}\end{pmatrix}\,,\\
 B_{tt\phi\phi}^I&=\begin{pmatrix} \delta_{i_1 j_1}\delta_{i_2 j_2}\delta_{k_1 l_1}\\  
\delta_{i_1 k_1}\delta_{i_2 j_2}\delta_{j_1 l_1}\\ 
 \delta_{i_1 l_1}\delta_{i_2 j_2}\delta_{k_1 j_1}\end{pmatrix}\,,\\
 B_{ssss}&=1\,,\\
 B_{\phi s\phi s}&=\delta_{ik}\,,\\
 B_{t st s}&=\delta_{i_1k_1}\delta_{i_2k_2}\,,\\
 B_{ttss}&=\delta_{i_1j_1}\delta_{i_2j_2}\,,\\
 B_{\phi \phi ss}&=\delta_{ij}\,,\\
B_{\phi s\phi t}&=\delta_{il_1}\delta_{kl_2}\,,\\
B_{\phi \phi s t}&=\delta_{il_1}\delta_{jl_2}\,,\\
 B_{sttt}&=\delta_{j_1k_1}\delta_{j_2l_1}\delta_{k_2l_2}\,,\\
}
where the indices for each of the four operators are labeled as $i,j,k,l$ respectively, all indices with the same letter should be symmetrized with the trace removed, and for simplicity we suppress the indices on the left-hand side. For the first four configurations with non-trivial bases, we can find the tensor structure using the rank-2 \SOthree Casimir $C$ acting on a basis $B$ with $n,m$ number of $i,j$ indices, respectively, as:
\es{casimir}{
CB^I=\sum_{J}M^{IJ}B^J\,,\qquad C\equiv(G_{i_1'}^{i_1}\oplus \dots \oplus G_{i_n'}^{i_n}\oplus G_{j_1'}^{j_1}\oplus \dots \oplus G_{j_m'}^{j_m})^2\,,
}
where $G$ are the usual \SOthree generators. The $K$ eigenvectors $(T_K)^J$ of $M^{IJ}$ are eigenvectors of $C$:
\es{eigs}{
(CT_K)^I=\sum_J M^{IJ}(T_K)^J=c_K(T_K)^I\, ,
}
where the eigenvalue $c$ for a rank $q$ \SOthree irrep is $q(q+1)$, which allows us to identify each $T_K$ with an irrep. Up to an overall normalization, these $T_K$ are then the desired tensor structures. For the last 8 configurations there is only one basis element, so the tensor structure is simply that element also up to an overall normalization. The final list of tensor structures is then
\es{Tfinal}{
\langle \phi\phi\phi\phi\rangle&:\qquad T^{\bf0^+}_{\bf1^-\bf1^-\bf1^-\bf1^- }=B^1_{\phi\phi\phi\phi}\,,\\ &\qquad\;\;\; T^{\bf1^+}_{\bf1^-\bf1^-\bf1^-\bf1^-}=B^2_{\phi\phi\phi\phi}-B^3_{\phi\phi\phi\phi}\,,\\
&\qquad\;\;\; T^{\bf2^+}_{\bf1^-\bf1^-\bf1^-\bf1^-}=B^2_{\phi\phi\phi\phi}+B^3_{\phi\phi\phi\phi}-\frac23B^1_{\phi\phi\phi\phi}\,,\\
\langle tttt\rangle&:\qquad T^{\bf0^+}_{\bf2^+\bf2^+\bf2^+\bf2^+}=B^1_{tttt}\,,\\
&\qquad\;\;\; T^{\bf1^+}_{\bf2^+\bf2^+\bf2^+\bf2^+}=B^4_{tttt}-B^5_{tttt}\,,\\
&\qquad\;\;\; T^{\bf2^+}_{\bf2^+\bf2^+\bf2^+\bf2^+}=B^4_{tttt}+B^5_{tttt}-\frac23B^1_{tttt}\,,\\
&\qquad\;\;\; T^{\bf3^+}_{\bf2^+\bf2^+\bf2^+\bf2^+}=B^5_{tttt}-B^4_{tttt}-\frac54B^3_{tttt}+\frac54B^2_{tttt}\,,\\
&\qquad\;\;\; T^{\bf4^+}_{\bf2^+\bf2^+\bf2^+\bf2^+}= -B^5_{tttt}-B^4_{tttt}+\frac{7}{12}B^3_{tttt}+\frac{7}{12}B^2_{tttt}+\frac{13}{30}B^1_{tttt}\,,\\
\langle t\phi t \phi\rangle,\langle \phi tt \phi\rangle&:\qquad T^{\bf1^-}_{\bf2^+\bf1^-\bf2^+\bf1^-}=2B^2_{t\phi t\phi}\,,\\
&\qquad\;\;\; T^{\bf2^-}_{\bf2^+\bf1^-\bf2^+\bf1^-}=-4B^1_{t\phi t\phi}+2B^2_{t\phi t\phi}+4B^3_{t\phi t\phi}\,,\\
&\qquad\;\;\; T^{\bf3^-}_{\bf2^+\bf1^-\bf2^+\bf1^-}=10B^1_{t\phi t\phi}-8B^2_{t\phi t\phi}+20B^3_{t\phi t\phi}\,,\\
\langle t t \phi\phi\rangle&:\qquad T^{\bf0^+}_{\bf2^+\bf2^+\bf1^-\bf1^-}=B^1_{t t\phi\phi}\,,\\
&\qquad\;\;\; T^{\bf1^+}_{\bf2^+\bf2^+\bf1^-\bf1^-}=B^2_{tt\phi \phi}-B^3_{t t\phi\phi}\,,\\
&\qquad\;\;\; T^{\bf2^+}_{\bf2^+\bf2^+\bf1^-\bf1^-}=B^2_{tt\phi \phi}+B^3_{t t\phi\phi}-\frac23B^1_{tt\phi \phi}\,,\\
\langle ssss\rangle&:\qquad T^{\bf0^+}_{\bf0^+\bf0^+\bf0^+\bf0^+}=B_{ssss}\,,\\
\langle \phi s\phi s\rangle,\langle  s\phi\phi s\rangle&:\qquad T^{\bf1^-}_{\bf1^-\bf0^+\bf1^-\bf2^+}=B_{\phi s\phi s}\,,\\
\langle tsts\rangle,\langle stts\rangle&:\qquad T^{\bf2^+}_{\bf2^+\bf0^+\bf2^+\bf0^+}=B_{t st s}\,,\\
\langle ttss\rangle&:\qquad T^{\bf0^+}_{\bf2^+\bf2^+\bf0^+\bf0^+}=B_{ttss}\,,\\
\langle \phi\phi ss\rangle&:\qquad T^{\bf0^+}_{\bf1^-\bf1^-\bf0^+\bf0^+}=B_{\phi \phi ss}\,,\\
\langle \phi s\phi t\rangle,\langle  s\phi\phi t\rangle&:\qquad T^{\bf1^-}_{\bf1^-\bf0^+\bf1^-\bf2^+}={\sqrt{2}}B_{\phi s\phi t}\,,\\
\langle \phi\phi st\rangle&:\qquad T^{\bf2^+}_{\bf1^-\bf1^-\bf0^+\bf2^+}=\sqrt{2}B_{\phi \phi s t}\,,\\
\langle sttt\rangle,\langle ttst\rangle&:\qquad T^{\bf2^+}_{\bf0^+\bf2^+\bf2^+\bf2^+}=\sqrt{2}B_{sttt}\,.\\
}
The overall normalization of these tensor structures has been chosen so that the OPE coefficients $\lambda_{\varphi_1\varphi_2 \cO}$ and $\lambda_{\varphi_3\varphi_4 \cO}$ in \eqref{4point} are consistent under permutation of their subscripts. This can be checked using the free theory, where we have the operators
\es{free}{
s(x)\equiv \frac{1}{\sqrt{6}}\phi^i(x)\phi^i(x)\,,\qquad t^{ij}(x)\equiv\frac{1}{\sqrt{2}}\phi^i(x)\phi^j(x) - \mathrm{trace}\,,
}
which have been normalized consistent with the 2-point function normalization in \eqref{2points}. We can then compute all the 4-point functions in table \ref{table} using Wick contractions and expand in blocks as in \eqref{4point} using the tensor structures in \eqref{Tfinal} to verify this consistency.\footnote{Note that there is another convention generated by the package \texttt{autoboot} \cite{Go:2019lke}. Our results for scanned external OPEs are different from the \texttt{autoboot} (ab) convention by
\begin{align}
\left( { {\lambda _{sss}^{{\rm{us}}}} , {\lambda _{tts}^{{\rm{us}}}} , {\lambda _{\phi \phi t}^{{\rm{us}}}} , {\lambda _{\phi \phi s}^{{\rm{us}}}} , {\lambda _{ttt}^{{\rm{us}}}} } \right) = \left( { {\lambda _{sss}^{{\rm{ab}}}} ,\frac{1}{{\sqrt 3 }} {\lambda _{tts}^{{\rm{ab}}}} ,\frac{1}{{\sqrt 5 }} {\lambda _{\phi \phi t}^{{\rm{ab}}}} ,\frac{1}{{\sqrt {10} }} {\lambda _{\phi \phi s}^{{\rm{ab}}}} ,\sqrt {\frac{6}{{35}}}  {\lambda _{ttt}^{{\rm{ab}}}} } \right)\,.
\end{align}
}

\section{Computed points}
\label{app:points}

In table~\ref{tab:allpoints} we list the 38 primal points we have computed in the $\Lambda=43$ island and in table~\ref{tab:disallowedpoints} we list the 270 dual points we computed at $\Lambda=43$. In table \ref{tab:the7points} we list the 7 primal points we use for the optimization computations described in section \ref{optimal}. 

\begin{figure}[htbp]
  \centering
  \includegraphics[width=0.85\textwidth]{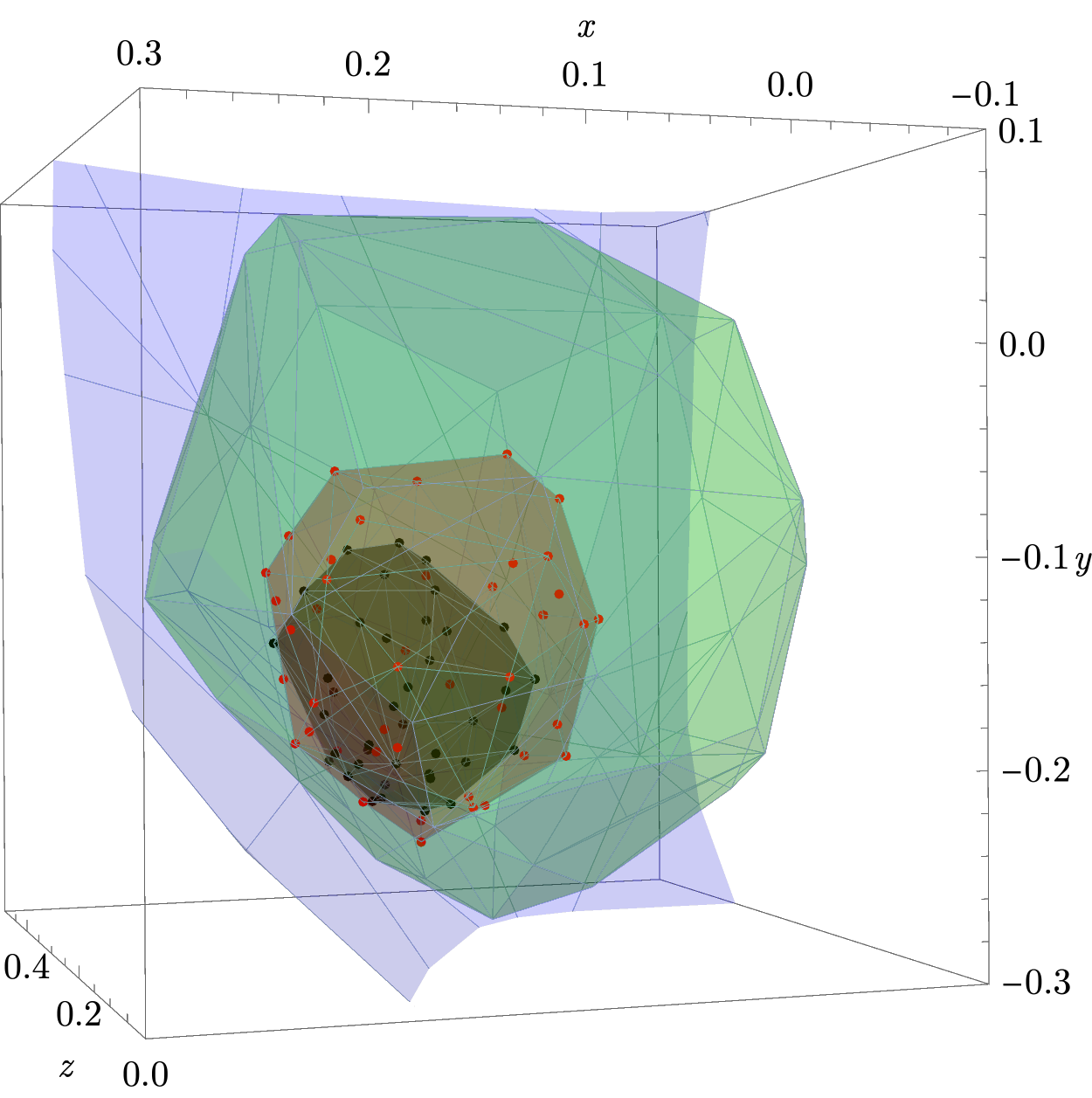}
  \caption{\label{dimension:affine} The convex hulls of the allowed points in the affine space (\ref{eq:affine}) at derivative orders $\Lambda=19,27,35,43$. The red and black data points are the allowed points at derivative orders $\Lambda=35,43$.}
\end{figure}

In figure \ref{dimension:affine} we show a plot of the allowed regions at $\Lambda=19,27,35,43$ after performing an affine transformation which makes the $\Lambda=19$ region roughly spherical. The precise affine transformation is given by:
\begin{align}
\label{eq:affine}
x &= 228.67 - 107.177 \Delta_{s} - 43.8661 \Delta_{t} - 8.77302 \Delta_{\f}\,,\nn\\
y &= -1061.39 - 694.406 \Delta_{s} + 1612.44 \Delta_{t} + 420.885 \Delta_{\f}\,,\\
z &= 2590.87 - 221.685 \Delta_{s} + 2629.52 \Delta_{t} - 10439.6 \Delta_{\f}\,.\nn
\end{align}

\begin{table}
\tiny
\centering
\begin{tabular}{@{}|c|c|c|c|c|c|c|@{}}
\toprule
$\De_\f$ & $\De_s$ & $\De_t$ & $\frac{{\lambda_{sss}}}{\lambda_{\phi \phi s}}$ & $\frac{{\lambda_{tts}}}{\lambda_{\phi \phi s}}$ & $\frac{{\lambda_{\phi \phi t}}}{\lambda_{\phi \phi s}}$ & $\frac{{\lambda_{t t  t}}}{\lambda_{\phi \phi s}}$\\
\midrule
0.5189783882& 1.5953612741& 1.2097311776& 0.9658557781& 1.8764272526& 1.6683150562& 2.8608280295\\
 0.5189583670& 1.5949959168& 1.2096121876& 0.9637866930& 1.8759071995& 1.6681321958& 2.8604010933\\
 0.5189461401& 1.5949711389& 1.2095536502& 0.9650503920& 1.8758868056& 1.6680723076& 2.8601976693\\
 0.5189272852& 1.5948074081& 1.2094888929& 0.9652812111& 1.8758487781& 1.6680235398& 2.8603061675\\
 0.5189613339& 1.5952564268& 1.2096662222& 0.9662461846& 1.8763712143& 1.6682528867& 2.8607475247\\
 0.5189198114& 1.5946719225& 1.2094285955& 0.9639220579& 1.8755694851& 1.6679344313& 2.8599409571\\
 0.5189172850& 1.5946165394& 1.2094312995& 0.9634877179& 1.8756194443& 1.6679524398& 2.8599835329\\
 0.5189500473& 1.5951798121& 1.2096307308& 0.9661780974& 1.8763286685& 1.6682360908& 2.8607267382\\
 0.5189649901& 1.5951958587& 1.2096821376& 0.9647455098& 1.8763265571& 1.6682560921& 2.8608385887\\
 0.5189431822& 1.5950799657& 1.2095793546& 0.9658559545& 1.8761787760& 1.6681550694& 2.8604066434\\
 0.5189526027& 1.5949370220& 1.2096034154& 0.9633651701& 1.8758925982& 1.6681462865& 2.8605568949\\
 0.5189301757& 1.5949315300& 1.2095370407& 0.9653175705& 1.8761152596& 1.6681181238& 2.8605816730\\
 0.5189168372& 1.5946844827& 1.2094296026& 0.9642949506& 1.8756631385& 1.6679573604& 2.8599418324\\
 0.5189483062& 1.5950509260& 1.2095711164& 0.9654214538& 1.8760347497& 1.6681115766& 2.8602582414\\
 0.5189153515& 1.5946825124& 1.2094352034& 0.9643765341& 1.8757126829& 1.6679642216& 2.8601068005\\
 0.5189440155& 1.5949715807& 1.2095885922& 0.9644538043& 1.8760784038& 1.6681817119& 2.8607365030\\
 0.5189150284& 1.5945099455& 1.2094139921& 0.9622684128& 1.8754680443& 1.6679176085& 2.8599290431\\
 0.5189347282& 1.5947841912& 1.2094914029& 0.9639498317& 1.8756547299& 1.6679896459& 2.8599829650\\
 0.5189802791& 1.5952998277& 1.2097251950& 0.9649312284& 1.8762516023& 1.6682831892& 2.8609105511\\
 0.5189248611& 1.5948368279& 1.2094835818& 0.9649249785& 1.8759055628& 1.6680304724& 2.8600899846\\
 0.5189306747& 1.5946954535& 1.2094799696& 0.9630209667& 1.8756016368& 1.6679818979& 2.8600556364\\
 0.5189217812& 1.5947698290& 1.2094894198& 0.9646493947& 1.8759369337& 1.6680732162& 2.8603682382\\
 0.5189121284& 1.5945494559& 1.2093932170& 0.9632128469& 1.8754599512& 1.6678971836& 2.8597295069\\
 0.5189738261& 1.5953116389& 1.2096932265& 0.9658322350& 1.8762663003& 1.6682491526& 2.8606277206\\
 0.5189145348& 1.5947341024& 1.2094532479& 0.9650247104& 1.8759430722& 1.6680183536& 2.8602402387\\
 0.5189014384& 1.5944048949& 1.2093555400& 0.9627737264& 1.8753951336& 1.6678520403& 2.8597952050\\
 0.5189305457& 1.5947604700& 1.2094652572& 0.9640748832& 1.8756103384& 1.6679623986& 2.8598809964\\
 0.5189623990& 1.5950949062& 1.2096629593& 0.9640070854& 1.8761554772& 1.6682313428& 2.8608189553\\
 0.5189460789& 1.5950582403& 1.2095808918& 0.9654253734& 1.8761261589& 1.6681362224& 2.8605045135\\
 0.5189301505& 1.5949452685& 1.2095278745& 0.9657542568& 1.8760854806& 1.6681017321& 2.8604962327\\
 0.5189685635& 1.5953320391& 1.2097096529& 0.9660028358& 1.8764583565& 1.6683135549& 2.8609575117\\
 0.5189497511& 1.5949209137& 1.2095606097& 0.9638109344& 1.8758202300& 1.6680711351& 2.8601899822\\
 0.5189337664& 1.5947155621& 1.2095107201& 0.9623655424& 1.8757184969& 1.6680386790& 2.8602450223\\
 0.5189453714& 1.5951023898& 1.2096147301& 0.9659065454& 1.8763651079& 1.6682206940& 2.8607919521\\
 0.5189862601& 1.5954028918& 1.2097515683& 0.9654844905& 1.8763339257& 1.6683105873& 2.8607881451\\
 0.5189476979& 1.5949533896& 1.2096010827& 0.9637175659& 1.8760304271& 1.6681861098& 2.8606729585\\
 0.5189346015& 1.5950348003& 1.2095678602& 0.9659534043& 1.8763062992& 1.6681684361& 2.8606248190\\
 0.5189598320& 1.5951484795& 1.2096664021& 0.9647455098& 1.8763265571& 1.6682560921& 2.8608385887\\
\bottomrule
\end{tabular}
\caption{\label{tab:allpoints}Allowed points in the $\Lambda=43$ island.}
\end{table}

\begin{table}
\tiny
\centering
\begin{tabular}{@{}|c|c|c|@{}}
\toprule
$\De_\f$ & $\De_s$ & $\De_t$ \\
\midrule
0.5187966798& 1.5921850401& 1.2086228804\\
 0.5189025751& 1.5940361780& 1.2092959408\\
 0.5189114364& 1.5946504985& 1.2093872462\\
 0.5189189599& 1.5941920882& 1.2093299741\\
 0.5189288736& 1.5938535802& 1.2091078506\\
 0.5186945881& 1.5924442671& 1.2084245865\\
 0.5187373587& 1.5921586352& 1.2084312743\\
 0.5187762022& 1.5918820333& 1.2086546805\\
 0.5187826937& 1.5914255278& 1.2085310641\\
 0.5188353864& 1.5934222782& 1.2090020249\\
 0.5188362222& 1.5935429810& 1.2089818386\\
 0.5188416799& 1.5933781659& 1.2089481869\\
 0.5188478354& 1.5938857812& 1.2091362879\\
 0.5188535569& 1.5932852248& 1.2088214042\\
 0.5188541349& 1.5932916366& 1.2089652272\\
 0.5188547707& 1.5939362783& 1.2090312421\\
 0.5188571369& 1.5930390946& 1.2090569647\\
 0.5188574085& 1.5927203987& 1.2089917699\\
 0.5188612075& 1.5939110280& 1.2091455740\\
 0.5188615055& 1.5937167137& 1.2090909898\\
 0.5188768144& 1.5938159784& 1.2090216235\\
 0.5188830465& 1.5941282784& 1.2092443533\\
 0.5188846453& 1.5942465723& 1.2092589387\\
 0.5188859294& 1.5941864473& 1.2092761662\\
 0.5188862012& 1.5941115729& 1.2092255065\\
 0.5188873828& 1.5941812791& 1.2092506161\\
 0.5188874108& 1.5939452740& 1.2091859961\\
 0.5188875107& 1.5943066473& 1.2093007022\\
 0.5188888052& 1.5938613363& 1.2092121233\\
 0.5188955343& 1.5940894591& 1.2092127567\\
 0.5188965815& 1.5941441764& 1.2092975081\\
 0.5189023610& 1.5945285422& 1.2093834019\\
 0.5189038211& 1.5934538031& 1.2090700892\\
 0.5189054986& 1.5948072619& 1.2094576026\\
 0.5189068088& 1.5944337603& 1.2092909633\\
 0.5189083086& 1.5946728229& 1.2094131489\\
 0.5189105727& 1.5946476632& 1.2094486445\\
 0.5189114164& 1.5934097919& 1.2091414652\\
 0.5189153608& 1.5942042741& 1.2092982492\\
 0.5189179786& 1.5948735383& 1.2094980056\\
 0.5189191601& 1.5945565549& 1.2094640368\\
 0.5189241090& 1.5941509824& 1.2094002783\\
 0.5189333103& 1.5939599259& 1.2094523297\\
 0.5189333930& 1.5943501521& 1.2094232463\\
 0.5189346097& 1.5944851028& 1.2095740716\\
 0.5189237305& 1.5944279251& 1.2093905391\\
 0.5188043639& 1.5926034786& 1.2087905625\\
 0.5188819788& 1.5940503654& 1.2091664050\\
 0.5189348544& 1.5939659673& 1.2093605905\\
 0.5189331311& 1.5945586909& 1.2094861373\\
 0.5188609431& 1.5938359757& 1.2091138392\\
 0.5189041783& 1.5931200581& 1.2091359461\\
 0.5189211553& 1.5940068208& 1.2093711976\\
 0.5188230608& 1.5930667503& 1.2087833097\\
 0.5187918880& 1.5920291631& 1.2086556637\\
 0.5187954081& 1.5933857238& 1.2088722712\\
 0.5188120708& 1.5921939042& 1.2085475622\\
 0.5188685720& 1.5940780889& 1.2092200213\\
 0.5188737587& 1.5938867482& 1.2091282102\\
 0.5188791085& 1.5939475441& 1.2092269843\\
 0.5188860122& 1.5944086593& 1.2093118142\\
 0.5188881256& 1.5943425103& 1.2092606077\\
 0.5188962192& 1.5944000610& 1.2093036926\\
 0.5188995836& 1.5944714851& 1.2093479909\\
 0.5189042967& 1.5942997732& 1.2093450874\\
 0.5189061160& 1.5944245329& 1.2093054811\\
 0.5189130711& 1.5943416289& 1.2092938887\\
 0.5189282439& 1.5948696505& 1.2094354495\\
 0.5189381636& 1.5945869176& 1.2095373290\\
 0.5189209582& 1.5946145623& 1.2093971026\\
 0.5188528255& 1.5932931016& 1.2088647649\\
 0.5188758367& 1.5937600844& 1.2090694206\\
 0.5189034036& 1.5943173907& 1.2092753786\\
 0.5187933545& 1.5923524886& 1.2085851052\\
 0.5188028340& 1.5925358527& 1.2086181925\\
 0.5188672706& 1.5938859690& 1.2090784902\\
 0.5188721453& 1.5939744974& 1.2092114019\\
 0.5188815521& 1.5940429270& 1.2090230925\\
 0.5188876803& 1.5938597752& 1.2092439737\\
 0.5188893843& 1.5942285412& 1.2092430003\\
 0.5188959297& 1.5932555635& 1.2092613868\\
 0.5189185378& 1.5944533142& 1.2093899362\\
 0.5189206686& 1.5947871150& 1.2094529310\\
 0.5189513139& 1.5946586522& 1.2094523424\\
 0.5189385685& 1.5946849443& 1.2094603545\\
 0.5189444575& 1.5946106034& 1.2094426944\\
 0.5189597959& 1.5951033035& 1.2095684829\\
 0.5190271291& 1.5953552765& 1.2099320047\\
 0.5191718508& 1.5977020194& 1.2105467890\\
 0.5189415400& 1.5950976756& 1.2095440010\\
\bottomrule
\end{tabular}
\begin{tabular}{@{}|c|c|c|@{}}
\toprule
$\De_\f$ & $\De_s$ & $\De_t$ \\
\midrule
0.5189424979& 1.5947703013& 1.2093601723\\
 0.5189461102& 1.5944802637& 1.2095417129\\
 0.5189461142& 1.5952505758& 1.2096113811\\
 0.5189509729& 1.5951396676& 1.2096608087\\
 0.5189524992& 1.5951035671& 1.2095733650\\
 0.5189561892& 1.5950657026& 1.2096530941\\
 0.5189575751& 1.5955030599& 1.2096401327\\
 0.5189576562& 1.5952623581& 1.2096851550\\
 0.5189595475& 1.5954833608& 1.2097148970\\
 0.5189682325& 1.5951956784& 1.2096321472\\
 0.5189696503& 1.5942534718& 1.2093617878\\
 0.5189733687& 1.5949010519& 1.2095786060\\
 0.5189772029& 1.5950953535& 1.2096440413\\
 0.5189800482& 1.5949290496& 1.2097035540\\
 0.5189804744& 1.5953372481& 1.2097875328\\
 0.5189818184& 1.5960483373& 1.2098711894\\
 0.5189830613& 1.5952764767& 1.2098175729\\
 0.5189871624& 1.5963425034& 1.2098915261\\
 0.5189966010& 1.5956243058& 1.2098589501\\
 0.5190066292& 1.5959897652& 1.2099481084\\
 0.5190086343& 1.5954797940& 1.2098872690\\
 0.5190093925& 1.5948495201& 1.2095311073\\
 0.5190140405& 1.5954232716& 1.2098887955\\
 0.5190158019& 1.5958546312& 1.2098800029\\
 0.5190176624& 1.5957396035& 1.2099524800\\
 0.5190182636& 1.5952950369& 1.2097197630\\
 0.5190222887& 1.5962382824& 1.2100406599\\
 0.5190294335& 1.5944921267& 1.2097386508\\
 0.5190368914& 1.5960448579& 1.2099540543\\
 0.5190379431& 1.5958717009& 1.2099655030\\
 0.5190482917& 1.5959534075& 1.2099669869\\
 0.5190587928& 1.5959983074& 1.2100169645\\
 0.5190598740& 1.5957892478& 1.2100080216\\
 0.5190647361& 1.5962663027& 1.2101521345\\
 0.5190695509& 1.5960823090& 1.2101350323\\
 0.5190725277& 1.5960088046& 1.2100991203\\
 0.5190732657& 1.5962258325& 1.2100744510\\
 0.5190787856& 1.5961544869& 1.2101199950\\
 0.5190839387& 1.5962931060& 1.2101834489\\
 0.5190846538& 1.5964564464& 1.2101823824\\
 0.5190882951& 1.5952923328& 1.2099488238\\
 0.5191014483& 1.5962764902& 1.2102433612\\
 0.5191200121& 1.5963878404& 1.2104254692\\
 0.5191238375& 1.5961605267& 1.2100960457\\
 0.5191427139& 1.5977670021& 1.2106878967\\
 0.5191456549& 1.5971361619& 1.2105308642\\
 0.5191695884& 1.5970766016& 1.2106595054\\
 0.5191747564& 1.5961581015& 1.2104165938\\
 0.5191752118& 1.5973305922& 1.2104827329\\
 0.5191772163& 1.5979425702& 1.2107834542\\
 0.5191998406& 1.5976763974& 1.2108343370\\
 0.5192110760& 1.5974046747& 1.2105784428\\
 0.5192313582& 1.5976397828& 1.2109134929\\
 0.5192577054& 1.5985007420& 1.2110602270\\
 0.5192609179& 1.5973050662& 1.2108724994\\
 0.5192751871& 1.5986516951& 1.2111693213\\
 0.5193182009& 1.5988546582& 1.2112832624\\
 0.5193313395& 1.5984303509& 1.2111533312\\
 0.5193414850& 1.5980461969& 1.2111569121\\
 0.5193551983& 1.5998541317& 1.2115697595\\
 0.5193848130& 1.5991038253& 1.2114973135\\
 0.5193945892& 1.5989679837& 1.2114217221\\
 0.5190798632& 1.5964346557& 1.2102061791\\
 0.5189741917& 1.5955685395& 1.2097697715\\
 0.5189927891& 1.5951952177& 1.2097372791\\
 0.5189996154& 1.5950788378& 1.2097054113\\
 0.5190167890& 1.5955602397& 1.2098873787\\
 0.5190308122& 1.5949198011& 1.2096389844\\
 0.5190339170& 1.5955996031& 1.2098749137\\
 0.5190412371& 1.5962115376& 1.2100681124\\
 0.5191177942& 1.5966893312& 1.2103147962\\
 0.5192445379& 1.5985977149& 1.2110282728\\
 0.5192695593& 1.5991098729& 1.2112586408\\
 0.5193036599& 1.5986614215& 1.2113279731\\
 0.5192414561& 1.5971357017& 1.2106354701\\
 0.5189651659& 1.5950121240& 1.2096667988\\
 0.5191751265& 1.5980350400& 1.2107387866\\
 0.5189584563& 1.5953669776& 1.2096954753\\
 0.5189653470& 1.5950297999& 1.2096904832\\
 0.5189699612& 1.5955018495& 1.2097342211\\
 0.5189780238& 1.5954734077& 1.2097071812\\
 0.5189801701& 1.5956399491& 1.2098195524\\
 0.5189846429& 1.5956222286& 1.2098043739\\
 0.5189848116& 1.5942072297& 1.2094316101\\
 0.5189955714& 1.5956904175& 1.2098601902\\
 0.5189970055& 1.5958004005& 1.2098315960\\
 0.5190050996& 1.5957883004& 1.2098865696\\
 0.5190163339& 1.5957550559& 1.2099183538\\
 0.5190199557& 1.5956651380& 1.2099295296\\
 0.5190607004& 1.5962908080& 1.2100873837\\
\bottomrule
\end{tabular}
\begin{tabular}{@{}|c|c|c|@{}}
\toprule
$\De_\f$ & $\De_s$ & $\De_t$ \\
\midrule
0.5190663597& 1.5958743309& 1.2100957055\\
 0.5190718073& 1.5963854342& 1.2101543103\\
 0.5191015148& 1.5964779979& 1.2103173913\\
 0.5191201873& 1.5971879432& 1.2103557329\\
 0.5191327281& 1.5966479322& 1.2102157357\\
 0.5192390608& 1.5990575125& 1.2111165018\\
 0.5189384637& 1.5945786868& 1.2092964612\\
 0.5189388374& 1.5949295375& 1.2094886518\\
 0.5189406470& 1.5951736959& 1.2096025578\\
 0.5189435236& 1.5946628177& 1.2095079568\\
 0.5189591857& 1.5937291370& 1.2094479043\\
 0.5189690300& 1.5947953102& 1.2095700139\\
 0.5189782490& 1.5955014573& 1.2097394518\\
 0.5189785861& 1.5948000056& 1.2096217602\\
 0.5189943897& 1.5951836165& 1.2096726879\\
 0.5190020000& 1.5955721180& 1.2098614948\\
 0.5190031045& 1.5956425756& 1.2098293707\\
 0.5190089551& 1.5959677551& 1.2099413822\\
 0.5190264203& 1.5960422185& 1.2100190836\\
 0.5190349642& 1.5958180205& 1.2099706558\\
 0.5190621203& 1.5957337543& 1.2098389282\\
 0.5191180161& 1.5968444953& 1.2103456280\\
 0.5191470052& 1.5962315204& 1.2104571613\\
 0.5191921249& 1.5963777247& 1.2105975545\\
 0.5192666256& 1.5975999519& 1.2109769054\\
 0.5192834446& 1.5981834807& 1.2109478678\\
 0.5192153931& 1.5971409678& 1.2108619123\\
 0.5190126129& 1.5960328832& 1.2098411820\\
 0.5189956968& 1.5952697384& 1.2097891345\\
 0.5190907117& 1.5964405597& 1.2102564468\\
 0.5189972378& 1.5954623091& 1.2097288249\\
 0.5192342602& 1.5977739085& 1.2106788412\\
 0.5192967458& 1.5978374167& 1.2109654958\\
 0.5190531962& 1.5957547434& 1.2100253487\\
 0.5190597017& 1.5959435763& 1.2100960808\\
 0.5189393710& 1.5943081218& 1.2094518082\\
 0.5189538314& 1.5947940605& 1.2094896280\\
 0.5189547388& 1.5948394180& 1.2095585036\\
 0.5189639253& 1.5945186785& 1.2094725381\\
 0.5189836057& 1.5957551038& 1.2098219526\\
 0.5189905121& 1.5953498417& 1.2098007194\\
 0.5189919017& 1.5941900169& 1.2095288310\\
 0.5189979105& 1.5956955072& 1.2097796040\\
 0.5189995667& 1.5957402879& 1.2099058216\\
 0.5190123754& 1.5941042932& 1.2095390773\\
 0.5190157406& 1.5958569876& 1.2099200545\\
 0.5190278431& 1.5965081852& 1.2100281015\\
 0.5190325131& 1.5959081051& 1.2099696538\\
 0.5190520384& 1.5963152404& 1.2100841277\\
 0.5190588609& 1.5960509826& 1.2099090548\\
 0.5190715439& 1.5955912797& 1.2101497526\\
 0.5190926525& 1.5965090465& 1.2102403519\\
 0.5191170800& 1.5974916154& 1.2105064335\\
 0.5191778676& 1.5977308111& 1.2107843923\\
 0.5192544590& 1.5971285489& 1.2107746440\\
 0.5192666631& 1.5983714488& 1.2109356823\\
 0.5192812420& 1.5987160882& 1.2111101581\\
 0.5193040795& 1.5983604351& 1.2112010479\\
 0.5189685870& 1.5947440353& 1.2095546973\\
 0.5189469362& 1.5948003211& 1.2094569896\\
 0.5190500327& 1.5959140301& 1.2100673844\\
 0.5190853457& 1.5968203648& 1.2103356646\\
 0.5189402432& 1.5947339426& 1.2095132572\\
 0.5189866072& 1.5952602047& 1.2097386461\\
 0.5189466850& 1.5949712041& 1.2095439276\\
 0.5189977538& 1.5955819151& 1.2098187686\\
 0.5189417109& 1.5948151822& 1.2095116242\\
 0.5189887395& 1.5955311410& 1.2097998738\\
 0.5189805152& 1.5954860043& 1.2097574571\\
 0.5189068037& 1.5944354150& 1.2093631340\\
 0.5189664963& 1.5953900793& 1.2097088870\\
 0.5189046136& 1.5944739467& 1.2093498390\\
 0.5189477115& 1.5951943364& 1.2096143706\\
 0.5189476015& 1.5951048660& 1.2095857135\\
 0.5189392465& 1.5947339121& 1.2095257880\\
 0.5189758138& 1.5951810417& 1.2097131902\\
 0.5189281878& 1.5946813688& 1.2094996850\\
 0.5189741332& 1.5953607762& 1.2097417847\\
 0.5189225607& 1.5945597787& 1.2094380766\\
 0.5189250556& 1.5947901155& 1.2094599723\\
 0.5189931813& 1.5954181855& 1.2097690833\\
 0.5189417323& 1.5950007532& 1.2095363269\\
 0.5189436065& 1.5950358588& 1.2096059939\\
 0.5189964571& 1.5954554607& 1.2098005401\\
 0.5189717084& 1.5953355582& 1.2096832796\\
 0.5189194848& 1.5945704303& 1.2094129825\\
 0.5189261393& 1.5947676104& 1.2095065854\\
 0.5189163665& 1.5945874291& 1.2093899278\\
 0.5188888691& 1.5942500939& 1.2092859143\\
 0.5189931888& 1.5954995182& 1.2097983737\\
\bottomrule
\end{tabular}
\caption{\label{tab:disallowedpoints}Disallowed points computed at $\Lambda=43$.}
\end{table}

\begin{table}
\tiny
\centering
\begin{tabular}{@{}|c|c|c|c|c|c|c|c|c|@{}}
\toprule
$\De_\f$ & $\De_s$ & $\De_t$ & $\frac{{\lambda_{sss}}}{\lambda_{\phi \phi s}}$ & $\frac{{\lambda_{tts}}}{\lambda_{\phi \phi s}}$ & $\frac{{\lambda_{\phi \phi t}}}{\lambda_{\phi \phi s}}$ & $\frac{{\lambda_{t t  t}}}{\lambda_{\phi \phi s}}$  & $\Delta_{T'}-\Delta_{T}$& $\Delta_{J'} - \Delta_{J}$\\
\midrule
0.5189121284&1.594549456&1.209393217&0.9632128469&1.875459951&1.667897184&2.859729507 & 1.0 & 0.1\\
0.5189145348&1.594734102&1.209453248&0.9650247103&1.875943072&1.668018354&2.860240239 & 1.0 & 1.0\\
0.5189337664&1.594715562&1.209510720&0.9623655424&1.875718497&1.668038679&2.860245022 & 1.0 & 1.0\\
0.5189415373&1.594941048&1.209557043&0.9646177852&1.875978443&1.668107872&2.860397307 & 1.0 & 1.0\\
0.5189431822&1.595079966&1.209579355&0.9658559545&1.876178776&1.668155069&2.860406643 & 1.0 & 0.1\\
0.5189685635&1.595332039&1.209709653&0.9660028358&1.876458357&1.668313555&2.860957512 & 1.0 & 0.1\\
0.5189862601&1.595402892&1.209751568&0.9654844905&1.876333926&1.668310587&2.860788145 & 1.0 & 0.1\\
\bottomrule
\end{tabular}
\caption{\label{tab:the7points}Allowed points in the $\Lambda=43$ island used to obtain bounds on $\lambda_{\f\f s}$, $C_T$, and $C_J$, along with the gaps above $\Delta_T$ and $\Delta_J$ that were assumed.}
\end{table}

\bibliography{Biblio}
\bibliographystyle{utphys}
\end{document}